\documentclass[aps,final,onecolumn,superscriptaddress,a4paper,showkeys]{revtex4}
\usepackage[tbtags]{amsmath}
\usepackage{amssymb}
\usepackage{enumerate}
\usepackage{graphicx}
\newcommand{\eg}{e.g.}
\newcommand{\ie}{i.e.}
\newcommand{\Tr}{\mathop{\text{Tr}}}
\newcommand{\sgn}{\mathop{\text{sgn}}}

\renewcommand{\Im}{\mathop{\text{Im}}}
\newcommand{\pri}{^{\prime}}
\newcommand{\pripri}{^{\prime\prime}}
\newcommand{\omb}{\omega_{\text{B}}}
\newcommand{\omc}{\omega_{\text{c}}}
\newcommand{\I}{\text{i}}
\renewcommand{\d}{\text{d}}
\newcommand{\e}{\text{e}}
\begin{document}
%
%
\title{Quantum Brownian motion in ratchet potentials: duality relation and its consequences}
%
%
\author{J. Peguiron}
\altaffiliation[Present address:]{Department of Physics and Astronomy, University of Basel, Klingelbergstrasse 82, CH-4056 Basel, Switzerland.}
\affiliation{Kavli Institute of Nanoscience, Delft University of Technology, Lorentzweg 1, 2628 CJ Delft, The
Netherlands} \affiliation{Institut f\"ur Theoretische Physik, Universit\"at Regensburg, D-93040 Regensburg,
Germany}
\author{M. Grifoni}
\affiliation{Institut f\"ur Theoretische Physik, Universit\"at
Regensburg, D-93040 Regensburg, Germany}
\date{November 18, 2005}
\begin{abstract}
Quantum Brownian motion in ratchet potentials is investigated by means of an approach based on a duality
relation. This relation links the long-time dynamics in a tilted ratchet potential in the presence of
dissipation with the one in a driven dissipative tight-binding model. The application to quantum ratchets yields
a simple expression for the ratchet current in terms of the transition rates in the tight-binding system.
\end{abstract}
\maketitle
%
%
Brownian motion in ratchet potentials~\cite{RatRev} has attracted a lot of interest. One reason is the fact that
ratchet systems, \ie\ periodic structures with broken spatial symmetry, present the property of allowing
transport under the influence of unbiased forces. The interest has recently grown with the transfer of the
problem in the quantum regime. There, the description of dissipative tunneling~\cite{WeiBK99} presents a
theoretical challenge which was tackled in relatively few
works~\cite{ReiPRL97,RonPRL98,GriPRL02,LehPRL02,SchPRB02,MacPRE04}, while first experimental realizations were
reported~\cite{LinSci99,MajPRL03}. After the semiclassical work~\cite{ReiPRL97}, further progress towards the
theoretical description of quantum ratchets involved modeling in terms of tight-binding
systems~\cite{RonPRL98,GriPRL02} or in terms of a molecular wire~\cite{LehPRL02}. Other available methods
include perturbation theory~\cite{SchPRB02} or a quantum Smoluchowski equation~\cite{MacPRE04}. Most of these
methods~\cite{ReiPRL97,RonPRL98,GriPRL02,MacPRE04} are restricted to the regime of moderate-to-strong friction.

The approach discussed in this article originates from works~\cite{SchPRL83,GuiPRL85,FisPRB85,ZwePRB87,SasPRB96}
on quantum Brownian motion in a tilted sinusoidal potential, which led to a duality relation for the mobility of
the system considered with the one of a driven dissipative tight-binding model. This idea emerged first in
Ref.~\cite{SchPRL83}, where the linear dc mobility at zero temperature was considered. In that work the interest
was focused on the occurrence of a diffusion-to-localization transition in the system with increasing
dissipation. These results were corroborated by means of renormalization-group methods~\cite{GuiPRL85}. The
duality relation was extended to the nonlinear dc mobility at finite temperatures in Ref.~\cite{FisPRB85}, where
an extensive physical discussion as well as important milestones of the proof were given. It was subsequently
applied to the investigation of the current-voltage characteristic of small Josephson junctions~\cite{ZwePRB87}.
Later, an identical duality relation for the linear ac mobility was obtained by a different approach in the
frame of linear response~\cite{SasPRB96}. In particular, that work went beyond the case of a strictly Ohmic
dissipative bath considered in Ref.~\cite{FisPRB85}, and included the case of an Ohmic bath with finite cutoff
frequency as well as sub and super-Ohmic baths. In Ref.~\cite{PegPRE05}, the formalism of Ref.~\cite{FisPRB85}
was generalized to arbitrary ratchet potentials, \ie, periodic potentials of arbitrary shape. Moreover, the
duality relation was extended to the average position of the quantum particle at long time. In the present
article, we give detailed proofs and discussion of these last results. The duality relation is obtained in its
most general form, in terms of the generating function out of which the average position, the mobility, and
other dynamical quantities can be extracted. Its domain of validity includes weak dissipation and nonlinear
driving. However, our demonstration remains restricted to the case of a strictly Ohmic bath.

The article is structured as follows: The problem is defined in Section~\ref{Sec:SP}; The duality relations and
their discussion are presented in Section~\ref{Sec:RD}; The application to the evaluation of the ratchet current
is developed in Section~\ref{Sec:App}; The details of the proofs are found in Section~\ref{Sec:P}.
\section{Statement of the problem}\label{Sec:SP}
We consider the Hamiltonian $\hat{H}_\text{R}$ of a quantum particle of mass $M$ in a one-dimensional periodic
potential $V(q)$ tilted by a force $F$,
\begin{equation}\label{Eq:DefHR}
\hat{H}_{\text{R}}=\frac{\hat{p}^2}{2M}+V(\hat{q})-F\hat{q}.
\end{equation}
The potential can be any function of periodicity~$L$, and is fully characterized by the amplitudes~$V_{l}$ and
phases~$\varphi_{l}$ of its harmonics in the Fourier representation
\begin{equation}\label{Eq:DefPot}
V(\hat{q})=\sum_{l=1}^{\infty}{V_{l}\cos{\left(2\pi l\hat{q}/L-\varphi_{l}\right)}}.
\end{equation}
In order to investigate quantum Brownian motion, we have to let the particle interact with a dissipative thermal
environment. This is modeled by the standard Hamiltonian~$\hat{H}_{\text{B}}$ of a bath of harmonic oscillators
whose coordinates are bilinearly coupled to the system coordinate~$\hat{q}$~\cite{WeiBK99}
\begin{equation}
\hat{H}_{\text{B}}=\frac{1}{2}\sum_{\alpha=1}^N\left[\frac{\hat{p}_\alpha^2}{m_\alpha}+m_\alpha\omega_\alpha^2\left(\hat{x}_\alpha-\frac{c_\alpha}{m_\alpha\omega_\alpha^2}\hat{q}\right)^2\right].
\end{equation}
The bath is fully characterized by its spectral density $J(\omega)=(\pi/2)\sum_{\alpha=1}^N(
c_\alpha^2/m_\alpha\omega_\alpha){\delta(\omega-\omega_\alpha)}$, defined in terms of the masses~$m_\alpha$,
frequencies~$\omega_\alpha$, and coupling strengths~$c_\alpha$ of the oscillators. We consider an Ohmic spectral
density, \ie\ linear~$J(\omega)\sim\eta\omega$ at low frequency~$\omega$. The viscosity coefficient~$\eta$,
together with the particle mass~$M$, defines the time scale of dissipation~$\gamma^{-1}=(\eta/M)^{-1}$.

The information on the system dynamics is contained in the reduced density matrix
$\hat{\rho}(t)=\Tr_{\text{B}}\hat{W}(t)$, obtained from the density matrix~$\hat{W}(t)$ of the system-plus-bath
$\hat{H}=\hat{H}_{\text{R}}+\hat{H}_{\text{B}}$ by performing the trace over the bath degrees of freedom. The
diagonal elements~$P(q,t)=\langle q|\hat{\rho}(t)|q\rangle$ of the reduced density matrix suffice in order to
evaluate, \eg, the evolution of the average position
$\langle\hat{q}(t)\rangle=\Tr_{\text{R}}\lbrace\hat{q}\hat{\rho}(t)\rbrace$. It turns out to be very powerful to
work with the generating function defined as
\begin{equation}\label{Eq:DefGF}
\tilde{P}(\lambda,t)=\int\d q\e^{\lambda q}P(q,t).
\end{equation}
The normalization of the reduced density matrix implies the property $\tilde{P}(\lambda=0,t)=1$. The name and
utility of the generating function come from the fact that its derivatives generate expectation values of powers
of the position operator
\begin{equation}\label{Eq:MomentsPos}
\frac{\partial^{k}}{\partial\lambda^k}\tilde{P}(\lambda,t)\biggr|_{\lambda=0}=\langle\hat{q}^k(t)\rangle.
\end{equation}

In particular, we would like to investigate quantum Brownian motion in a ratchet system. Apart from special
configurations $\lbrace V_{l}\sin{(\varphi_{l}-l\varphi_{1})}=0\ \forall l\rbrace$ of the amplitudes~$V_{l}$ and
phases~$\varphi_{l}$, the potential~(\ref{Eq:DefPot}) is spatially asymmetric and describes such a ratchet
system. The ratchet effect is characterized by a nonvanishing average stationary particle current
$v_\text{R}^{\infty} =\lim_{t\to\infty}t^{-1}\int_{0}^{t}\d t^{\prime}v(t^{\prime})$ in the presence of unbiased
time-dependent driving~$F(t\pri)$, characterized by $\lim_{t\to\infty}t^{-1}\int_{0}^{t}\d
t^{\prime}F(t^{\prime})=0$, and switched on at time $t^{\prime}=t_0$. In this article, we report a method to
evaluate the stationary velocity~$v^{\infty}_\text{DC}(F)$, which is simply obtained by time-differentiation
of~$\langle\hat{q}(t)\rangle$, in the biased situation~(\ref{Eq:DefHR}) of time-independent driving~$F$. The
ratchet current in the presence of unbiased bistable driving switching adiabatically between the values~$\pm F$
is obtained through the relation $v_\text{R}^\infty= v^{\infty}_\text{DC}(F)+v^{\infty}_\text{DC}(-F)$.

The diagonal elements~$P(q,t)$ of the reduced density matrix can be obtained by real-time path integrals
techniques~\cite{WeiBK99,KleBK04}. At initial time~$t^{\prime}=t_0$, we assume a preparation in a product form
 where the bath is in thermal equilibrium with the system $\hat{W}(t_0)=\hat{\rho}(t_0)\e^{-\beta\hat{H}_{\text{B}}^{(0)}}
\lbrack\Tr_\text{B}\e^{-\beta\hat{H}_{\text{B}}^{(0)}}\rbrack^{-1}$. In~$\hat{H}_{\text{B}}^{(0)}$, the system
operator~$\hat{q}$ is replaced by an initial position~$q_0$. The bath temperature is fixed by $T=1/\beta
k_{\text{B}}$. This leads to the expression
\begin{equation}
P(q,t)=\langle q|\hat{\rho}(t)|q\rangle=\int\d q_i \int\d q_i\pri\langle q_i|\hat{\rho}(t_0)|q_i\pri\rangle
G(q,q,q_i,q_i\pri,t)
\end{equation}
with the propagating function
\begin{equation}\label{Eq:DefPF}
G(q_f,q_f\pri,q_i,q_i\pri,t)=\int_{q_i}^{q_f}\mathcal{D}q\int_{q_i\pri}^{q_f\pri}\mathcal{D}^*q\pri
A[q,t]A^*[q\pri,t]F_{\text{FV}}[q,q\pri,t]
\end{equation}
given as a double path integral on the paths $q(t\pri)$ and $q\pri(t\pri)$. For the continuous coordinate $q$
and a Hamiltonian of the form~(\ref{Eq:DefHR}), the path integral stands for \cite[ch.~2]{KleBK04}
\begin{equation}\label{Eq:DefPI}
\int_{q_i}^{q_f}\mathcal{D}q=\lim_{N_\text{I}\to\infty}\left(\frac{M}{2\pi\I\hbar\Delta\tau}\right)^{N_\text{I}/2}\int\d q_1\int\d
q_2\ldots\int\d q_{N_\text{I}-1},
\end{equation}
where the time interval $t-t_0$ has been sliced in~$N_\text{I}$ intervals of length $\Delta\tau=(t-t_0)/N_\text{I}$ and the path
$q(t\pri)$ has been discretized into the set of values $q_k=q(t_0+k\Delta\tau)$ for $k=1,\ldots,N_\text{I}-1$. The
boundaries of the path integral remind of $q(t_0)=q_i$ and $q(t)=q_f$. The propagator $A[q,t]$ reads
\begin{equation}
A[q,t]=\exp\left\lbrace\frac{\I}{\hbar}\int_{t_0}^t\d
t\pri\left[\frac{M}{2}\dot{q}(t\pri)^2-V(q(t\pri))+Fq(t\pri)\right]\right\rbrace.
\end{equation}
The Feynman-Vernon influence functional may conveniently be written as
$F_{\text{FV}}[q,q\pri,t]=\exp\lbrace\Phi_{\text{FV}}[\xi,\chi,t]\rbrace$, in terms of the phase
\begin{multline}\label{Eq:DefPhiFV}
\Phi_{\text{FV}}[\xi,\chi,t]=-\frac{1}{\hbar}\int_{t_0}^{t}\d t\pri\int_{t_0}^{t\pri}\d t\pripri\xi(t\pri)L_{\text{R}}(t\pri-t\pripri)\xi(t\pripri)-\frac{2\I}{\hbar}\int_{t_0}^{t}\d t\pri\int_{t_0}^{t\pri}\d t\pripri\xi(t\pri)L_{\text{I}}(t\pri-t\pripri)\chi(t\pripri)\\
-\frac{2\I}{\hbar}M_{\text{I}}(0)\int_{t_0}^{t}\d
t\pri\xi(t\pri)\chi(t\pri)+\frac{2\I}{\hbar}q(t_0)\int_{t_0}^{t}\d t\pri\xi(t\pri)M_{\text{I}}(t\pri-t_0),
\end{multline}
which depends on the difference~$\xi(t\pri)=q(t\pri)-q\pri(t\pri)$ and
average~$\chi(t\pri)=[q(t\pri)+q\pri(t\pri)]/2$ of the two paths. It induces nonlocal-in-time Gaussian
correlations between the paths. The functions $L_{\text{R}}(\tau)$ and $L_{\text{I}}(\tau)$ denote the real and
imaginary part of the bath correlation function
\begin{equation}
L(\tau)=\frac{1}{\pi}\int_0^{\infty}\d\omega
J(\omega)\left[\coth\left(\frac{\hbar\omega\beta}{2}\right)\cos(\omega\tau)-\I\sin(\omega\tau)\right].
\end{equation}
Integrating the imaginary part yields the function~$M_{\text{I}}(\tau)=\int_0^{\infty}\d\omega
J(\omega)\cos(\omega\tau)/\pi\omega$.

Our goal is to evaluate the generating function~(\ref{Eq:DefGF}). In general, the nonlinearity of the
potential~$V(q)$ prevents the evaluation of the path integrals. However, an exact expansion of the contribution
of the potential in the propagator~$A\lbrack q\rbrack$ makes it feasible. In the following, we shall present
this method, which was first introduced by Fisher and Zwerger~\cite{FisPRB85} for a sinusoidal potential.
\section{Results and Discussion}\label{Sec:RD}
In order to expand the contribution of the potential, we introduce a ``charge''~$\sigma$ taking values in the
set $\lbrace\pm 1, \pm2, \ldots\rbrace$, and corresponding amplitudes defined as
\begin{align}\label{Eq:DefDelta}
\Delta_{\sigma}&=\frac{V_{\sigma}}{2}\e^{\I\varphi_{\sigma}}\quad\text{for }\sigma>0,
&\Delta_{-\sigma}&=\Delta_{\sigma}^*.
\end{align}
This allows to rewrite the potential~(\ref{Eq:DefPot}) as a simple sum of exponentials
\begin{equation}\label{Eq:PotSumExp}
V(q)=\sum_{\sigma=\pm 1,\pm 2,\ldots}\Delta_{\sigma}\e^{-2\pi\I\sigma q/L}.
\end{equation}
Using this expansion, one can demonstrate that~(see Section~\ref{SubSec:ExpPot})
\begin{equation}\label{Eq:ExpPot}
\exp\left\lbrace-\frac{\I}{\hbar}\int_{t_0}^t\d t\pri
V(q(t\pri))\right\rbrace=\sum_{n=0}^{\infty}\sum_{\lbrace\sigma_j\rbrace}\prod_{j=1}^n\left(-\frac{\I\Delta_{\sigma_j}}{\hbar}\right)
\int_{t_0}^t\d t_n\int_{t_0}^{t_n}\d t_{n-1}\ldots\int_{t_0}^{t_2}\d
t_1\exp\left\lbrace-\frac{\I}{\hbar}\int_{t_0}^t\d t\pri\rho(t\pri)q(t\pri)\right\rbrace,
\end{equation}
where we have introduced $n$~charges~$\sigma_j$ and corresponding times~$t_j$, as well as the function
$\rho(t\pri)=(2\pi\hbar/L)\sum_{j=1}^{n}\sigma_j\delta(t\pri-t_j)$. This expression can be substituted in the
propagator~$A[q,t]$. We have to do the same for the second propagator~$A^*[q\pri,t]$, using the complex
conjugate of Eq.~(\ref{Eq:ExpPot}) with a new set of $n\pri$~charges~$\sigma\pri_{j\pri}$ and corresponding
times~$t\pri_{j\pri}$, and a new function $\rho\pri(t\pri)$. The product~$A[q,t]A^*[q\pri,t]$ may then be
conveniently rewritten in terms of the difference~$\xi(t\pri)$, respectively average path~$\chi(t\pri)$,
\begin{equation}\label{Eq:ProdProp}
A[q,t]A^*[q\pri,t]=\boxed{\sum}\exp\biggl\lbrace\frac{\I}{\hbar}\int_{t_0}^t\d
t\pri\biggl[M\dot{\xi}(t\pri)\dot{\chi}(t\pri)-\chi(t\pri)\left[\rho(t\pri)-\rho\pri(t\pri)\right]-\xi(t\pri)\frac{1}{2}\left[\rho(t\pri)+\rho\pri(t\pri)\right]+F\xi(t\pri)\biggr]\biggr\rbrace.
\end{equation}
The gain of this expansion is that the paths now enter at most quadratically in the argument of the exponential.
Eventually, the path integrals will become Gaussian integrals. The price paid is the emergence of a series
expression. We have introduced a compact notation for the sums, products and integrals involved
\begin{equation}
\boxed{\sum}=
\sum_{n=0}^{\infty}\sum_{n\pri=0}^{\infty}\sum_{\lbrace\sigma_j\rbrace}\sum_{\lbrace\sigma\pri_{j\pri}\rbrace}\prod_{j=1}^n\left(-\frac{\I\Delta_{\sigma_j}}{\hbar}\right)
\prod_{j\pri=1}^{n\pri}\left(\frac{\I\Delta^*_{\sigma\pri_{j\pri}}}{\hbar}\right)
\int_{t_0}^t\d t_n\int_{t_0}^{t_n}\d t_{n-1}\ldots\int_{t_0}^{t_2}\d t_1\int_{t_0}^t\d
t\pri_{n\pri}\int_{t_0}^{t\pri_{n\pri}}\d t\pri_{n\pri-1}\ldots\int_{t_0}^{t\pri_2}\d t\pri_1.
\end{equation}
Performing the now Gaussian path integrals, and after a long calculation described in
Sections~\ref{SubSec:EvalPI}--\ref{SubSec:ExtractLambda}, we obtain our main result
\begin{multline}\label{Eq:DRGF}
\tilde{P}(\lambda,t)\sim\boxed{\sideset{}{^\prime}\sum}\Tr\nolimits_\text{R}\left\lbrace\hat{\rho}(t_0)\e^{\lambda\left(\hat{q}+\hat{p}/\eta\right)}\right\rbrace\exp\biggl\lbrace\Phi_{\text{FV}}^{\text{TB}}[y_{\text{sh}},x_{\text{sh}},t]+\frac{\I
F}{\hbar}\int_{t_0}^{t}\d t\pri y_{\text{sh}}(t\pri)\\
+\lambda\left[-\Delta\chi+\frac{F(t-t_0)}{\eta}+\frac{2\I k_{\text{B}}T}{\hbar}\int_{t_0}^{t}\d t\pri
y_{\text{sh}}(t\pri)\right]+\frac{\hbar\lambda^2}{2\eta^2}\left[N(t)-\frac{1}{\gamma}\dot{N}(t_0)-\I\eta\right]\biggr\rbrace.
\end{multline}
As indicated by the relation symbol, this result is valid within some approximations, as shown in the full
derivation of Section~\ref{Sec:P}. The validity regime of~(\ref{Eq:DRGF}) is discussed in detail below
Eq.~(\ref{Eq:DRPos}). The functions $y_{\text{sh}}(t\pri)$ and $x_{\text{sh}}(t\pri)$ denote the difference
$y_{\text{sh}}(t\pri)=q_{\text{sh}}(t\pri)-q\pri_{\text{sh}}(t\pri)$, respectively average
$x_{\text{sh}}(t\pri)=[q_{\text{sh}}(t\pri)+q\pri_{\text{sh}}(t\pri)]/2$, of the step-like paths
\begin{subequations}\label{Eq:DefTBPaths}
\begin{align}
\label{Eq:DefTBForwardPath}q_{\text{sh}}(t\pri)&=\tilde{L}\sum_{j=1}^n\sigma_j\left[\theta(t\pri-t_j)-1\right]+q_{\text{sh}}(t)\\
\label{Eq:DefTBBackwardPath}q\pri_{\text{sh}}(t\pri)&=\tilde{L}\sum_{j\pri=1}^{n\pri}\sigma\pri_{j\pri}\left[\theta(t\pri-t\pri_{j\pri})-1\right]+q_{\text{sh}}(t).
\end{align}
\end{subequations}
The step heights are multiples of~$\tilde{L}=2\pi\hbar/\eta L$. The paths end up at the same
value~$q_{\text{sh}}(t)$ at the final time~$t$. Therefore the difference path ends up at $y_{\text{sh}}(t)=0$.
The quantities
\begin{subequations}\label{Eq:DefDeltaxichi}
\begin{align}
\Delta\xi&=y_{\text{sh}}(t)-y_{\text{sh}}(t_0)=\tilde{L}\left[\sum\nolimits_{j=1}^n\sigma_j-\sum\nolimits_{j\pri=1}^{n\pri}\sigma\pri_{j\pri}\right]\\
\Delta\chi&=x_{\text{sh}}(t)-x_{\text{sh}}(t_0)=\frac{\tilde{L}}{2}\left[\sum\nolimits_{j=1}^n\sigma_j+\sum\nolimits_{j\pri=1}^{n\pri}\sigma\pri_{j\pri}\right]
\end{align}
\end{subequations}
specify the initial boundary condition for the two paths. The primed sum
\begin{equation}\label{Eq:DefPBsum}
\boxed{\sideset{}{^\prime}\sum}=\boxed{\sum}\ \delta\left(\frac{\Delta\xi}{\tilde{L}},0\right),
\end{equation}
where~$\delta$ denotes the Kronecker symbol, is thus restricted to the configurations for which the difference
path starts at~$y_{\text{sh}}(t_0)=0$. The influence
phase~$\Phi_{\text{FV}}^{\text{TB}}[y_{\text{sh}},x_{\text{sh}},t]$ is defined as in (\ref{Eq:DefPhiFV}),
provided that the spectral density~$J(\omega)$ entering the correlation functions $L(\tau)$ and
$M_\text{I}(\tau)$ is replaced by the new spectral density
\begin{equation}\label{Eq:DefSDTB}
J_{\text{TB}}(\omega)=\frac{J(\omega)}{1+(\omega/\gamma)^2}.
\end{equation}
Finally, the auxiliary function~$N(\tau)$ is discussed in Section~\ref{SubSec:ExtractLambda}.

The justification of these notations appears when one considers the generating
function~$\tilde{P}_\text{TB}(\lambda,t)$ of a driven tight-binding model given by the Hamiltonian
\begin{equation}\label{Eq:DefHTB}
\hat{H}_{\text{TB}}=\sum_{m=1}^{\infty}\sum_{l=-\infty}^{\infty}\Bigl(\Delta_{m}|l+m\rangle\langle l|
+\Delta_{m}^{*}|l\rangle\langle l+m|\Bigr)-F\hat{q}_{\text{TB}}.
\end{equation}
The couplings~$\Delta_m$ are precisely the one introduced in~(\ref{Eq:DefDelta}) and involved in the boxed sum.
The spatial periodicity of this tight-binding model, which can be read in the position operator
$\hat{q}_{\text{TB}}=\tilde{L}\sum_{l=-\infty}^{\infty}l|l\rangle\langle l|$, is precisely the height
unit~$\tilde{L}$ of the steps of the paths~(\ref{Eq:DefTBPaths}). This tight-binding model is bilinearly coupled
to a different bath of harmonic oscillators characterized by the spectral density~$J_\text{TB}(\omega)$ given
in~(\ref{Eq:DefSDTB}). This spectral density is still Ohmic, with the same viscosity coefficient~$\eta$, but now
presents a Drude cutoff at the frequency~$\gamma$ set by the dissipation of the original model. The system is
initially prepared in the state~$\hat{\rho}_\text{TB}(t_0)=|l_0\rangle\langle l_0|$ with
$l_0\tilde{L}=q_\text{sh}(t_0)$. In this situation, the generating function reads
\begin{equation}\label{Eq:GFTB}
\tilde{P}_\text{TB}(\lambda,t)=\boxed{\sideset{}{^\prime}\sum}\exp\left\lbrace\Phi_{\text{FV}}^{\text{TB}}[y_{\text{sh}},x_{\text{sh}},t]+\frac{\I
F}{\hbar}\int_{t_0}^{t}\d t\pri y_{\text{sh}}(t\pri)+\lambda\left(l_0\tilde{L}+\Delta\chi\right)\right\rbrace,
\end{equation}
which bears a clear structural resemblance with~(\ref{Eq:DRGF}). The $\lambda^2$-terms are absent
of~(\ref{Eq:GFTB}), but they do not play any role as far as one is interested in the average
position~$\langle\hat{q}(t)\rangle$~[see Eq.~(\ref{Eq:MomentsPos})]. One also notices that~$\Delta\chi$ comes
with an opposite sign in the two expressions.

The link between the original model~(\ref{Eq:DefHR}) and the tight-binding model~(\ref{Eq:DefHTB}) can be pushed
further. The normalization of the generating function $\tilde{P}(\lambda=0,t)=1$ yields the identity
\begin{equation}\label{Eq:IdPNorm}
1=\boxed{\sideset{}{^\prime}\sum}\exp\left\lbrace\Phi_{\text{FV}}^{\text{TB}}[y_{\text{sh}},x_{\text{sh}},t]+\frac{\I
F}{\hbar}\int_{t_0}^{t}\d t\pri y_{\text{sh}}(t\pri)\right\rbrace,
\end{equation}
starting either from (\ref{Eq:DRGF}) or (\ref{Eq:GFTB}). Differentiating with respect to~$F$ yields the set of
non-trivial identities
\begin{equation}\label{Eq:IdPNormdiffF}
0=\boxed{\sideset{}{^\prime}\sum}\left[\frac{\I}{\hbar}\int_{t_0}^{t}\d t\pri
y_{\text{sh}}(t\pri)\right]^k\exp\biggl\lbrace\Phi_{\text{FV}}^{\text{TB}}[y_{\text{sh}},x_{\text{sh}},t]+\frac{\I
F}{\hbar}\int_{t_0}^{t}\d t\pri y_{\text{sh}}(t\pri)\biggr\rbrace,
\end{equation}
for any~$k=1,2,\ldots,\infty$.

Another important result can be obtained from the relation~(\ref{Eq:DRGF}) by evaluating the average position
$\langle\hat{q}(t)\rangle=\left.[\partial\tilde{P}(\lambda,t)/\partial\lambda]\right|_{\lambda=0}$. Using the
identities (\ref{Eq:IdPNorm}) and (\ref{Eq:IdPNormdiffF}) with $k=1$, one gets
\begin{equation}\label{Eq:DRPos}
\langle\hat{q}(t)\rangle\sim\langle\hat{q}(t_0)\rangle+\frac{\langle\hat{p}(t_0)\rangle}{\eta}+\frac{F(t-t_0)}{\eta}-\langle\hat{q}_{\text{TB}}(t)\rangle_{\text{TB}},
\end{equation}
where $\langle\hat{q}(t_0)\rangle=\Tr\nolimits_\text{R}\left\lbrace\hat{q}\hat{\rho}(t_0)\right\rbrace$ and
$\langle\hat{p}(t_0)\rangle=\Tr\nolimits_\text{R}\left\lbrace\hat{p}\hat{\rho}(t_0)\right\rbrace$ denote the
position and momentum of the initial preparation of the ratchet system. The last term of~(\ref{Eq:DRPos}) is the
average of the position operator~$\hat{q}_{\text{TB}}$ in the driven dissipative tight-binding
model~(\ref{Eq:DefHTB}), initially prepared in the state~$\hat{\rho}_\text{TB}(t_0)=|0\rangle\langle 0|$. It can
be obtained from Eq.~(\ref{Eq:GFTB}) with
$\langle\hat{q}_\text{TB}(t)\rangle=\left.[\partial\tilde{P}_\text{TB}(\lambda,t)/\partial\lambda]\right|_{\lambda=0}$
and $l_0=0$. It comes with a minus sign due to the minus sign in front of~$\Delta\chi$ in~(\ref{Eq:DRGF}). The
duality relation for the position~(\ref{Eq:DRPos}) is a very useful result for quantum ratchet systems. We shall
discuss this application in Section~\ref{Sec:App}.

The duality relations (\ref{Eq:DRGF}) and (\ref{Eq:DRPos}) are approximate results, as denoted by the relation
symbol. As derived in this work, they are valid when the following conditions are simultaneously met:
\begin{enumerate}[i)]
\item Long-time dynamics: The measurement time~$t-t_0$ should be much longer than the time scale~$1/\gamma$
set by dissipation. This can be easily controlled experimentally.%
\item Rare transitions limit: The terms $\e^{-\gamma(t_\text{tr}-t_0)}$, $\e^{-\gamma(t-t_\text{tr})}$,
$\e^{-\omb(t_\text{tr}-t_0)}$, and $\e^{-\omb(t-t_\text{tr})}$, with~$\omb=2\pi k_\text{B}T/\hbar$, should be
negligible with respect to~$1$, when~$t_\text{tr}$ equals any of the times~$t_j,t\pri_{j\pri}$. These times,
which are integration variables involved in the boxed sum, are the transition times in the double path integral
representation~(\ref{Eq:GFTB}) of the generating function of the tight-binding model. Therefore, this
approximation corresponds to neglect, in the boxed sum, the contributions from the paths which involve
transitions on a time scale~$\max(\gamma^{-1},\omega_{\text{B}}^{-1})$ after the initial time~$t^{\prime}=t_0$ or
before the final time~$t^{\prime}=t$. It will therefore be valid when the transitions in the tight-binding model
are rare on a time scale~$\max(\gamma^{-1},\omega_{\text{B}}^{-1})$. This condition is controlled by the
dissipation and the
temperature.%
\end{enumerate}

Furthermore, in our derivation we have used a strictly Ohmic spectral density~$J(\omega)=\eta\omega$. In this
case, the function~$M_{\text{I}}(\tau)$ takes the simple form $M_{\text{I}}(\tau)=\eta\delta(\tau)$, and the
divergence of~$L_{\text{R}}^{\text{TB}}(0)$ allows to restrict the configuration sum to its primed version. A
physically more realistic situation would be to consider the Ohmic spectral
density~$J(\omega)=\eta\omega\e^{-\omega/\omc}$ with finite cutoff frequency~$\omc$.

We do not know to which extent these restrictions are specific to the method that we have used in order to
derive the duality relations. An equivalent duality relation for the mobilities~[see Eq.~(\ref{Eq:DRMobility})]
has been derived in Ref.~\cite{SasPRB96} in the frame of linear response for a sinusoidal potential. It is
interesting to notice that the derivation presented in that work does not require the restrictions to the rare
transitions limit and to a strictly Ohmic bath. A more generalized version of~(\ref{Eq:DRMobility}) has also
been obtained for a much broader class of spectral densities, including sub-Ohmic and super-Ohmic ones, and in
the case of time-dependent driving. We currently do not see any problem of principle in order to generalize our
demonstration for a general form of the spectral density, and this is the subject of work in progress. However,
we do not see how to avoid the restriction to the rare transitions limit in our derivation. Furthermore, we do
not know how to generalize the identities~(\ref{Eq:IdPNormdiffF}), which we have used in our proof, for the case
of time-dependent driving. These remain open questions.

Let us now give some interpretation of the results. The duality relation~(\ref{Eq:DRGF}) for the generating
function is not very useful in itself, but very powerful in order to generate useful results. It links the
dynamics of the two systems~(\ref{Eq:DefHR}) and~(\ref{Eq:DefHTB}). The precise relation between the two systems
is specified by: i) The relation~(\ref{Eq:DefDelta}) between the harmonics of the potential of the original
system and the couplings in the tight-binding system; ii) The relation~$\tilde{L}=2\pi\hbar/\eta L$ between the
spatial periodicities~$L$ of the original system, and~$\tilde{L}$ of the tight-binding system; iii) The
relation~(\ref{Eq:DefSDTB}) between the spectral densities of the baths of harmonics oscillators coupled to each
of the two systems. In the original system, the relevant dynamical parameters are captured by the dissipation
parameter~$\alpha=\eta L^2/2\pi\hbar$ and the energy drop per periodicity length~$\epsilon=FL$. Due to the
change of periodicity length, these parameters become~$\tilde{\alpha}=1/\alpha$ and~$\tilde{\epsilon}=\epsilon/{\alpha}$ in the tight-binding system. Thus, weak dissipation in one system maps to
strong dissipation in the other one, although the viscosity~$\eta$ in the spectral density does not change.

The duality relation for the average position~(\ref{Eq:DRPos}) is an example of a useful result which can be
extracted from~(\ref{Eq:DRGF}). There, the relation between the average positions $\langle\hat{q}(t)\rangle$ and
$\langle\hat{q}_{\text{TB}}(t)\rangle_{\text{TB}}$ in the two systems, which holds at long time, is explicit.
The asymptotic dynamics covered by this result is usually described in terms of the nonlinear mobility
$\mu=\lim_{t\to\infty}\langle\dot{\hat{q}}(t)\rangle/F$. Accordingly, the duality relation~(\ref{Eq:DRPos}) may
be rewritten in the form
\begin{equation}\label{Eq:DRMobility}
\mu(\alpha,\epsilon)\sim\mu_{0}-\mu_{\text{TB}}(1/\alpha,\epsilon/\alpha),
\end{equation}
where $\mu_{0}=1/\eta$ is the mobility of the free system, $V(\hat{q})\equiv 0$. In the special case of a
sinusoidal potential, this relation was already obtained in~\cite{FisPRB85} for the dc mobility.
As mentioned above, it has also been derived in~\cite{SasPRB96} for the linear ac mobility in a sinusoidal
potential.

The second derivative of the generating function~$\tilde{P}(\lambda,t)$ with respect to its parameter~$\lambda$
yields the variance~$\langle\hat{q}^2(t)\rangle$, which gives information about diffusion and current noise. It
would thus be natural to try to extract from~(\ref{Eq:DRGF}) a duality relation for this quantity. However, the
result diverges, because the quantity~$\dot{N}(t_0)$ involved in the right-hand side diverges for the strictly
Ohmic spectral density~$J(\omega)=\eta\omega$ considered in the derivation. In order to get results on diffusion
and current noise, we shall thus have to go beyond this approximation and allow for an Ohmic bath with finite
cutoff frequency. As mentioned above, we do not see any problem of principle in order to generalize our demonstration to
this situation.
\section{Application: Evaluation of the ratchet current}\label{Sec:App}
In this section, we shall discuss the application of~(\ref{Eq:DRPos}) to evaluate the current in ratchet
systems. By time-differentiation of~$\langle\hat{q}(t)\rangle$, given on the left-hand side of~(\ref{Eq:DRPos}),
one obtains the stationary velocity~$v^{\infty}_\text{DC}(F)$ in the biased situation of time-independent
driving~$F$. As discussed above, the ratchet current in the presence of unbiased bistable driving switching
adiabatically between the values~$\pm F$ is obtained through the relation $v_\text{R}^\infty=
v^{\infty}_\text{DC}(F)+v^{\infty}_\text{DC}(-F)$. Our task is therefore to evaluate the right-hand side
of~(\ref{Eq:DRPos}), in particular the average of the position operator~$\hat{q}_{\text{TB}}$ in the driven
dissipative tight-binding model~(\ref{Eq:DefHTB}), initially prepared in the
state~$\hat{\rho}_\text{TB}(t_0)=|0\rangle\langle 0|$. It can be obtained from the generating
function~$\tilde{P}_\text{TB}(\lambda,t)$ of the tight-binding system through
$\langle\hat{q}_\text{TB}(t)\rangle=\left.[\partial\tilde{P}_\text{TB}(\lambda,t)/\partial\lambda]\right|_{\lambda=0}$.

The generating function can be obtained from~(\ref{Eq:GFTB}) with~$l_0=0$. However, this formula is not the most
suitable in order to get the long-time behavior required in~(\ref{Eq:DRPos}). For this purpose, we go back to
the population~$P_l^\text{TB}(t)$ of the site~$|l\rangle$ of the tight-binding system at time~$t$, after a
preparation in the site~$|l_0\rangle=|0\rangle$ at initial time~$t_0$. These populations are the diagonal
elements of the reduced density matrix of the system~(\ref{Eq:DefHTB}) coupled to its bath of harmonic
oscillators of spectral density~$J_\text{TB}(\omega)$. They are related to the generating function by
$\tilde{P}_\text{TB}(\lambda,t)=\sum_{l=-\infty}^\infty\e^{\lambda l\tilde{L}}P_l^\text{TB}(t)$. The real-time
path integrals techniques which led to~(\ref{Eq:GFTB}) yield for the populations an expression which can be
rewritten in the form of an exact generalized master equation
\begin{equation}
\dot{P}_l^\text{TB}(t)=\int_{t_0}^t\d t\pri\sum_{l\pri=-\infty}^\infty
K_{l-l\pri}(t-t\pri)P_{l\pri}^\text{TB}(t\pri).
\end{equation}
For the rather technical discussion of the expressions for the kernels~$K_m(\tau)$, we refer to
Refs.~\cite{WeiBK99,GriPRE96}. The generalized master equation can be easily rewritten in terms of the
generating function, yielding
\begin{equation}\label{Eq:GME}
\dot{\tilde{P}}_\text{TB}(\lambda,t)=\int_{t_0}^t\d
t\pri\tilde{K}(\lambda,t-t\pri)\tilde{P}_\text{TB}(\lambda,t\pri),
\end{equation}
with $\tilde{K}(\lambda,\tau)=\sum_{m=-\infty}^\infty \e^{\lambda m\tilde{L}}K_m(\tau)$. As shown in
Section~\ref{SubSec:GFTBlongtime}, the solution of this equation behaves at long times as
\begin{equation}\label{Eq:GFlongtime}
\tilde{P}_\text{TB}(\lambda,t)\underset{t\to\infty}{\sim}\exp\left\lbrace(t-t_0)\sum_{m\ne0}\Gamma_m\left(\e^{\lambda
m\tilde{L}}-1\right)\right\rbrace.
\end{equation}
The transition rate~$\Gamma_m$ from a site~$|l\rangle$ to a site~$|l+m\rangle$ is obtained from the
corresponding kernel by $\Gamma_m=\int_0^\infty\d\tau K_m(\tau)$. From this result we derive easily
\begin{equation}
\langle\hat{q}_\text{TB}(t)\rangle\underset{t\to\infty}{\sim}(t-t_0)\tilde{L}\sum_{m\ne0}m\Gamma_m.
\end{equation}
Plugging this in the duality relation~(\ref{Eq:DRPos}) and differentiating, we obtain the stationary
velocity~$v^{\infty}_\text{DC}(F)$, respectively~$v^{\infty}_\text{DC}(-F)$, in terms of the transition
rates~$\Gamma_m(F)$, respectively~$\Gamma_m(-F)$,
\begin{equation}\label{Eq:Statv}
v_\text{DC}^\infty(\pm F)=\pm\frac{F}{\eta}-\tilde{L}\sum_{m\ne0}m\Gamma_m(\pm F).
\end{equation}
The ratchet current in the presence of adiabatic bistable driving reads accordingly
\begin{equation}\label{Eq:vR}
v_\text{R}^\infty=-\tilde{L}\sum_{m\ne0}m\left[\Gamma_m(F)+\Gamma_m(-F)\right].
\end{equation}
This result shows that the ratchet current in a system characterized by the potential~(\ref{Eq:DefPot}) is very
simply related to the transition rates in the tight-binding model~(\ref{Eq:DefHTB}). As the duality relation
from which it is derived, it is valid in the rare transitions limit $\Gamma_m(\pm F)\ll\min(\gamma,2\pi
k_\text{B}T/\hbar)$ and for a strictly Ohmic bath characterized by the spectral density~$J(\omega)=\eta\omega$.

The ultimate task is thus to evaluate the tight-binding transition rates~$\Gamma_m(\pm F)$. The expression
obtained by real-time path integrals techniques consists of numerous contributions which can be classified with
respect to the number~$N$ of transitions in the double tight-binding path which they involve. We denote
by~$\Gamma_m^{(N)}(\pm F)$ the sum of all contributions involving $N$~transitions and call it the $N$th order
transition rate. The total rate follows from $\Gamma_m(\pm F)=\sum_{N=2}^\infty\Gamma_m^{(N)}(\pm F)$. The
derivation of the explicit expressions are discussed in Section~\ref{SubSec:TBrates}.  At second order, we find,
for~$m\ne0$,
\begin{equation}\label{Eq:2ndOrderRate}
\Gamma_m^{(2)}(F)=\frac{\left|\Delta_m\right|^2}{\hbar^2}\int_{-\infty}^\infty\d\tau\e^{-m^2(\tilde{L}^2/\hbar)Q(\tau)+\I
m(F\tilde{L}/\hbar)\tau}.
\end{equation}
The twice-integrated bath correlation function is defined as
\begin{equation}\label{Eq:DefTIBathCorr}
Q(\tau)=\frac{1}{\pi}\int_0^{\infty}\d\omega
\frac{J_\text{TB}(\omega)}{\omega^2}\left[\coth\left(\frac{\hbar\omega\beta}{2}\right)\left[1-\cos(\omega\tau)\right]+\I\sin(\omega\tau)\right]
\end{equation}
with the spectral density~$J_\text{TB}(\omega)$ given in~(\ref{Eq:DefSDTB}). We remind that the
couplings~$\Delta_m$ are related to the harmonics of the potential~(\ref{Eq:DefPot}) through
Eq.~(\ref{Eq:DefDelta}). One sees that the phases~$\varphi_m$ of the potential, which are identified to the
phases of the couplings~$\Delta_m$, do not come into play at second order.

At third order, there are contributions which involve only $\Delta_{\pm1}$ and $\Delta_{\pm2}$. They read
\begin{equation}\label{Eq:3rdOrder112Rate}
\Gamma_m^{(3)}[112](F)=\frac{2\left|\Delta_1\right|^2\left|\Delta_2\right|}{\hbar^3}\Im\left\lbrace\int_{-\infty}^\infty\d\tau
G_{|m|}^{(3)}[112](\tau)\e^{\I m\frac{F\tilde{L}}{\hbar}\tau-\I\sgn(m)\varphi_{112}}\right\rbrace,
\end{equation}
for~$m=\pm1,\pm2$, with the functions
\begin{subequations}
\begin{align}
G_1^{(3)}[112](\tau)&=-\int_0^\infty\d\rho\e^{-\frac{2\tilde{L}^2}{\hbar}Q(-\rho)}
\Bigl[\e^{\frac{\tilde{L}^2}{\hbar}\left[-2Q(\tau+\rho)+Q(\tau+2\rho)\right]}+\e^{\frac{\tilde{L}^2}{\hbar}\left[-2Q(\tau-\rho)+Q(\tau-2\rho)\right]}\Bigr]
\\
G_2^{(3)}[112](\tau)&=\int_0^\infty\d\rho\e^{\frac{\tilde{L}^2}{\hbar}\left[Q(\rho)-2Q(\tau+\frac{1}{2}\rho)-2Q(\tau-\frac{1}{2}\rho)\right]}.
\end{align}
\end{subequations}
The phases~$\varphi_m$ of the potential enter through the unique phase difference
$\varphi_{112}=\varphi_2-2\varphi_1$.

There are also contributions involving $\Delta_{\pm1}$, $\Delta_{\pm2}$, and $\Delta_{\pm3}$. They may be
rewritten
\begin{equation}\label{Eq:3rdOrder123Rate}
\Gamma_m^{(3)}[123](F)=\frac{2\left|\Delta_1\right|\left|\Delta_2\right|\left|\Delta_3\right|}{\hbar^3}\Im\left\lbrace\int_{-\infty}^\infty\d\tau
G_{|m|}^{(3)}[123](\tau)\e^{\I m\frac{F\tilde{L}}{\hbar}\tau-\I\sgn(m)\varphi_{123}}\right\rbrace,
\end{equation}
for $m=\pm1,\pm2,\pm3$, with the functions
\begin{subequations}
\begin{align}
G_1^{(3)}[123](\tau)&=-\int_0^\infty\d\rho\e^{-\frac{6\tilde{L}^2}{\hbar}Q(-\rho)}
\Bigl[\e^{\frac{\tilde{L}^2}{\hbar}\left[-3Q(\tau+2\rho)+2Q(\tau+3\rho)\right]}+\e^{\frac{\tilde{L}^2}{\hbar}\left[-3Q(\tau-2\rho)+2Q(\tau-3\rho)\right]}\Bigr]
\\
G_2^{(3)}[123](\tau)&=-\int_0^\infty\d\rho\e^{-\frac{3\tilde{L}^2}{\hbar}Q(-\rho)}
\Bigl[\e^{\frac{\tilde{L}^2}{\hbar}\left[-6Q(\tau+\frac{1}{2}\rho)+2Q(\tau+\frac{3}{2}\rho)\right]}+\e^{\frac{\tilde{L}^2}{\hbar}\left[-6Q(\tau-\frac{1}{2}\rho)+2Q(\tau-\frac{3}{2}\rho)\right]}\Bigr]
\\
G_3^{(3)}[123](\tau)&=\int_0^\infty\d\rho\e^{\frac{2\tilde{L}^2}{\hbar}Q(\rho)}
\Bigl[\e^{\frac{\tilde{L}^2}{\hbar}\left[-6Q(\tau-\frac{1}{3}\rho)-3Q(\tau+\frac{2}{3}\rho)\right]}+\e^{\frac{\tilde{L}^2}{\hbar}\left[-6Q(\tau+\frac{1}{3}\rho)-3Q(\tau-\frac{2}{3}\rho)\right]}\Bigr],
\end{align}
\end{subequations}
and the phase difference
$\varphi_{123}=\varphi_3-\varphi_2-\varphi_1=(\varphi_3-3\varphi_1)-(\varphi_2-2\varphi_1)$.

If the potential~(\ref{Eq:DefPot}) sustains at most the first three harmonics, meaning that the
couplings~$\Delta_m$ are all~$0$ for~$|m|>3$, there are no other contributions to the third-order rates, which are
then given by
\begin{equation}
\Gamma_m^{(3)}(F)=\Gamma_m^{(3)}[112](F)+\Gamma_m^{(3)}[123](F).
\end{equation}

A very interesting feature of the expressions for the second-order and third-order rates is their explicit and
simple dependence on the moduli and phases of the couplings~$\Delta_m$, related to the amplitudes and phases of
the original potential~(\ref{Eq:DefPot}). In particular, this dependence has direct consequences for the ratchet
current~(\ref{Eq:vR}). Firstly, one can explicitly verify that the ratchet current vanishes for symmetric
potentials. The second-order rates satisfy $\Gamma_m^{(2)}(-F)=\Gamma_{-m}^{(2)}(F)$, therefore they do not
contribute to the ratchet current. Likewise, one can see that the third-order contributions to the ratchet
current, which are the dominant ones, are proportional to $\sin(\varphi_2-2\varphi_1)$ or
$\sin(\varphi_3-3\varphi_1)$. Therefore, they vanish for spatially symmetric potentials, characterized
by~$\lbrace V_{l}\sin{(\varphi_{l}-l\varphi_{1})}=0\ \forall l\rbrace$. Secondly, this simple dependence on the
potential parameters should be accessible in experimental realizations where the potential can be tailored, as,
\eg, in arrays of Josephson junctions~\cite{MajPRL03}.

We see that the complexity of the transition rates increases with the order~$N$. In tight-binding models with
large dissipation parameter~$\eta\tilde{L}^2/2\pi\hbar$ and/or high temperature, neglecting higher orders is
known to be a good approximation~\cite{WeiBK99}.

\begin{figure}
\begin{center}
\begin{tabular}{cc}
\includegraphics[width=7cm]{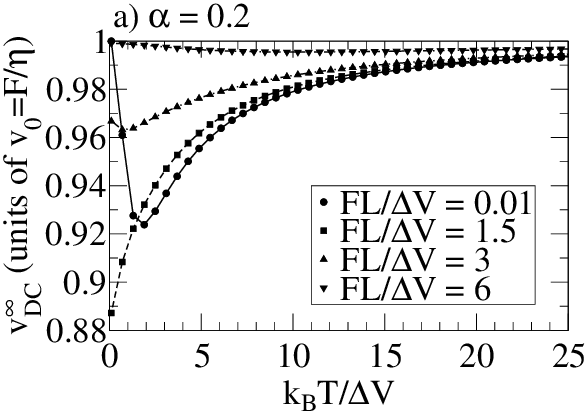}&
\includegraphics[width=7cm]{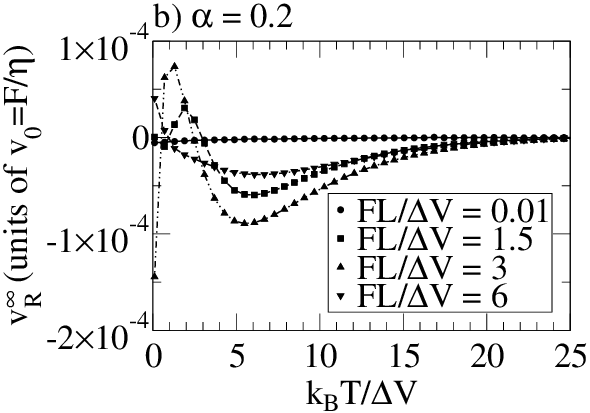}
\end{tabular}
\end{center}
\caption{Stationary velocity~(a) and ratchet current~(b) as a function of temperature, for different values of
the driving amplitude. Weak dissipation is chosen with~$\alpha=0.2$. For all curves, the point at the lowest
temperature is evaluated for~$k_\text{B}T/\Delta V=0.1$. The lines are guides for the eyes.}\label{Fig:1tmp}
\end{figure}
\begin{figure}
\begin{center}
\begin{tabular}{cc}
\includegraphics[width=7cm]{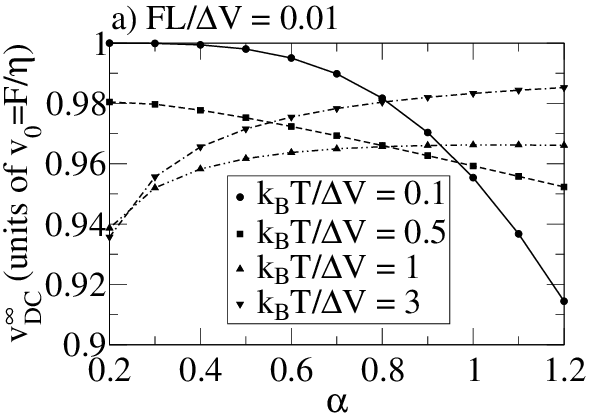}&
\includegraphics[width=7cm]{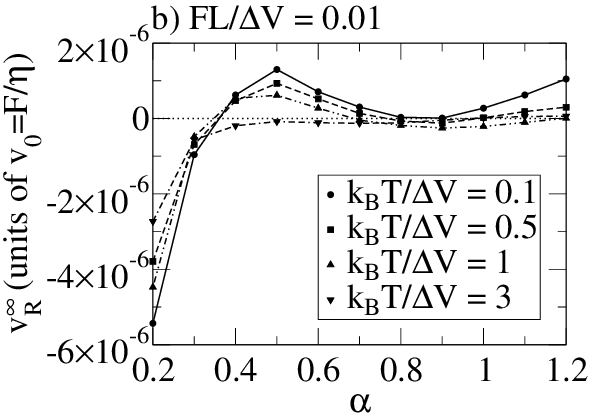}
\end{tabular}
\end{center}
\caption{Stationary velocity~(a) and ratchet current~(b) as a function of dissipation strength, for different
values of the temperature. Weak driving amplitude is chosen with~$FL/\Delta V=0.01$. In this regime, the
stationary velocity~(a) is a good approximation of the linear mobility, given in units of~$\mu_0=1/\eta$. The
lines are guides for the eyes.}\label{Fig:2diss}
\end{figure}

Finally, we discuss the behavior of the stationary velocity~(\ref{Eq:Statv}) and ratchet current~(\ref{Eq:vR})
as a function of the driving amplitude, the temperature, and the dissipation strength for a potential sustaining
two harmonics only. The rates are evaluated up to third order, using the expressions~(\ref{Eq:2ndOrderRate})
and~(\ref{Eq:3rdOrder112Rate}). The outcome is shown in Figs.~\ref{Fig:1tmp} and~\ref{Fig:2diss}. With~$V_1=4V_2$,
the untilted potential, obtained from~(\ref{Eq:DefPot}), has a barrier height~$\Delta V=2.2V_1$.
Another energy scale relevant for the dynamics is the confinement energy~$E_\text{cf}=2\pi\hbar^2/M L^2$ of a
particle of mass~$M$ confined within a length~$L$. The dynamical regime can be specified by the ratio between
these two quantities. With~$\Delta V/E_\text{cf}=0.264$, the typical action of the particle, $S=\sqrt{2M\Delta
VL^2}=\sqrt{\Delta V/E_\text{cf}\pi}\times h$, is about~$0.3 h$. The dissipation strength~$\alpha=\eta
L^2/2\pi\hbar$ is the ratio between the energy scale~$\hbar\gamma$ associated with the dissipation
rate~$\gamma=\eta/M$, and the confinement energy~$E_\text{cf}$. Another measure of the dissipation strength is
given by the ratio between~$\gamma$ and the classical oscillation frequency~$\Omega_0=2\pi\sqrt{V_1/ML^2}$ in
the untilted potential. With the value~$\alpha=0.2$ used in Fig.~\ref{Fig:1tmp}, this ratio,
$\gamma/\Omega_0=\alpha/\sqrt{2\pi V_1/E_\text{cf}}$, is about one fourth, which corresponds to weak
dissipation.

The behavior of the stationary velocity as a function of the temperature for different values of the driving
amplitude and weak dissipation~$\alpha=0.2$ is shown in Fig.~\ref{Fig:1tmp}a. The first curve at low
driving~$FL/\Delta V=0.01$~(circles) is also a good approximation of the linear mobility, given in units
of~$\mu_0=1/\eta$. The first value at low temperature~$k_\text{B}T/\Delta V=0.1$ is very close to the
velocity~$v_0=F/\eta$ obtained in the absence of the potential~$V\equiv0$. Upon increasing temperature, the
stationary velocity first decreases, before increasing again above a crossover temperature~$k_\text{B}T^*/\Delta
V\approx2$. This behavior is characteristic of weak dissipation. This crossover temperature is lower for the
third curve, at larger driving~$FL/\Delta V=3$~(up triangles). The ratchet current for the same parameter values
is shown in Fig.~\ref{Fig:1tmp}b. A non-monotonic behavior with reversals is found. From these two figures, one
also sees that the stationary velocity tends to~$v_0$ for driving amplitudes or temperatures much higher than
the potential barrier, and that the ratchet current vanishes correspondingly. A discussion of the behavior of
the stationary velocity and of the ratchet current as a function of the driving amplitude for the same system
can be found in Ref.~\cite{PegPRE05}.

The behavior of the stationary velocity as a function of the dissipation strength for different values of the
temperature and weak driving~$FL/\Delta V=0.01$ is shown in Fig.~\ref{Fig:2diss}a. At weak dissipation, the
stationary velocity~(and equivalently the linear mobility) tends to the value~$v_0$ obtained in the absence of
the potential upon decreasing the temperature, as already seen in Fig.~\ref{Fig:1tmp}a. This indicates
delocalization of the ground state. At strong dissipation, one finds an opposite behavior. The few curves shown
in the figure indicate that the transition between these different behaviors happens around~$\alpha=1$. The
analysis of the power-law dependence of the transition rates at low temperature shows that this transition
happens at~$\alpha=1$~\cite{PegPRE05}. The ratchet current, shown in Fig.~\ref{Fig:2diss}b for the same
parameter values, explicitly reveals the occurrence of reversals as a function of the dissipation strength.

In these numerical applications, none of the rates exceeds~$0.04\gamma$ and~$0.05\omega_{\text{B}}$, which means
that the duality relation is valid for this system. However, one should keep in mind that only contributions to
the tight-binding transition rates up to third order in tunneling have been evaluated. Higher-order
contributions could become relevant, \eg, at strong dissipation, when the dual tight-binding system experiences
weak dissipation~$\tilde{\alpha}=1/\alpha$.

\section{Proofs}\label{Sec:P}
\subsection{Expansion of the potential}\label{SubSec:ExpPot}
First, we shall demonstrate how the expansion~(\ref{Eq:PotSumExp}) of the potential leads to the series
expression~(\ref{Eq:ExpPot}). Using the power series representation of the exponential function, we write
\begin{equation}\label{Eq:ExpPowSer}
\exp\left\lbrace-\frac{\I}{\hbar}\int_{t_0}^t\d t\pri V(q(t\pri))\right\rbrace=\sum_{n=0}^{\infty}\int_{t_0}^t\d
t_n\int_{t_0}^{t_n}\d t_{n-1}\ldots\int_{t_0}^{t_2}\d t_1\prod_{j=1}^n\left[-\frac{\I}{\hbar}V(q(t_j))\right].
\end{equation}
Due to the complete symmetry of the integrand in all the $t_j$, the $n$ integrals have been entangled,
compensating the $1/n!$ factor coming from the series expansion. We now take advantage of the
representation~(\ref{Eq:PotSumExp}) of the potential as a sum, introducing a charge~$\sigma_j$ and
amplitudes~$\Delta_{\sigma_j}$ for each term~$V(q(t_j))$. By distributivity, product and sum can be exchanged,
yielding a sum on configurations $\sum_{\lbrace\sigma_j\rbrace}=\sum_{\sigma_1=\pm 1,\pm
2,\ldots}\ldots\sum_{\sigma_n=\pm 1,\pm 2,\ldots}$. This leads to
\begin{equation}
\prod_{j=1}^n\left[-\frac{\I}{\hbar}V(q(t_j))\right]=\sum_{\lbrace\sigma_j\rbrace}\prod_{j=1}^n\left(-\frac{\I\Delta_{\sigma_j}}{\hbar}\right)\exp\left\lbrace-\frac{2\pi\I}{L}\sum_{j=1}^n\sigma_j
q(t_j)\right\rbrace.
\end{equation}
With the help of the function~$\rho(t\pri)$ introduced below Eq.~(\ref{Eq:ExpPot}), this can be rewritten in
terms of the path~$q(t\pri)$
\begin{equation}
\prod_{j=1}^n\left[-\frac{\I}{\hbar}V(q(t_j))\right]=
\sum_{\lbrace\sigma_j\rbrace}\prod_{j=1}^n\left(-\frac{\I\Delta_{\sigma_j}}{\hbar}\right)\exp\left\lbrace-\frac{\I}{\hbar}\int_{t_0}^t\d
t\pri\rho(t\pri)q(t\pri)\right\rbrace.
\end{equation}
Substituting this expression in~(\ref{Eq:ExpPowSer}) demonstrates~(\ref{Eq:ExpPot}).
\subsection{Evaluation of the path integrals}\label{SubSec:EvalPI}
We now turn to the evaluation of the path integrals. We start from the expression for the propagating
function~(\ref{Eq:DefPF}). We rewrite the path integrals in terms of the difference~$\xi(t\pri)$, respectively
average path~$\chi(t\pri)$, with the boundary conditions $\xi_i=q_i-q\pri_i$, $\chi_i=(q_i+q\pri_i)/2$,
$\xi_f=q_f-q\pri_f$, and $\chi_f=(q_f+q\pri_f)/2$. Collecting the expressions (\ref{Eq:DefPhiFV}) and
(\ref{Eq:ProdProp}), we have
\begin{multline}
G(q_f,q_f\pri,q_i,q_i\pri,t)=\boxed{\sum}\int_{\xi_i}^{\xi_f}\mathcal{D}\xi\int_{\chi_i}^{\chi_f}\mathcal{D}^*\chi\exp\biggl\lbrace-S_{\text{R}}[\xi]-\I S_{\text{I}}[\xi]+\frac{\I F}{\hbar}\int_{t_0}^t\d t\pri\xi(t\pri)\\
+\frac{\I}{\hbar}\int_{t_0}^t\d t\pri\chi(t\pri)\left[-M\ddot{\xi}(t\pri)+\eta\dot{\xi}(t\pri)-\left[\rho(t\pri)-\rho\pri(t\pri)\right]\right]
+\frac{\I
M}{\hbar}\left[\dot{\xi}(t)\chi_f-\dot{\xi}(t_0)\chi_i\right]+\frac{\I\eta}{\hbar}\left[-\xi_f\chi_f+\xi_i\chi_i+\frac{1}{2}\xi_i^2\right]\biggr\rbrace,
\end{multline}
with the definitions
\begin{subequations}
\begin{align}
S_{\text{R}}[\xi]&=\frac{1}{\hbar}\int_{t_0}^{t}\d t\pri\int_{t_0}^{t\pri}\d
t\pripri\xi(t\pri)L_{\text{R}}(t\pri-t\pripri)\xi(t\pripri)\\
S_{\text{I}}[\xi]&=\frac{1}{\hbar}\int_{t_0}^t\d
t\pri\xi(t\pri)\frac{1}{2}\left[\rho(t\pri)+\rho\pri(t\pri)\right].
\end{align}
\end{subequations}
For the imaginary part of the bath correlation function, we have made use of the
simplification~$M_{\text{I}}(\tau)=\eta\delta(\tau)$ for a strictly Ohmic bath. We also have performed partial
integration in order to remove the dependence on the derivative~$\dot{\chi}(t\pri)$ of the average path.

We discretize the paths according to the procedure described below Eq.~(\ref{Eq:DefPI}). The integrals in the
argument of the exponential are discretized as in the following example
\begin{equation}
\int_{t_0}^t\d
t\pri\chi(t\pri)\left[-M\ddot{\xi}(t\pri)+\eta\dot{\xi}(t\pri)-\left[\rho(t\pri)-\rho\pri(t\pri)\right]\right]\\
=\sum_{k=0}^{N_\text{I}-1}\Delta\tau\chi_k\left[-M\ddot{\xi}_k+\eta\dot{\xi}_k-\left[\rho_k-\rho\pri_k\right]\right],
\end{equation}
and we evaluate the derivatives of the path $\xi$ with the difference formulae
$\dot{\xi}_k=(\xi_{k+1}-\xi_k)/\Delta\tau$ and $\ddot{\xi}_k=(\xi_{k+1}-2\xi_k+\xi_{k-1})/{\Delta\tau}^2$. The
terms involving $\chi_k$ form the integral
\begin{equation}
\int\d\chi_k\exp\left\lbrace\frac{\I}{\hbar}\chi_k\left[-M\ddot{\xi}_k+\eta\dot{\xi}_k-\left[\rho_k-\rho\pri_k\right]\right]\right\rbrace\\
=2\pi\delta\left(\frac{\I}{\hbar}\chi_k\left[-M\ddot{\xi}_k+\eta\dot{\xi}_k-\left[\rho_k-\rho\pri_k\right]\right]\right).
\end{equation}
This $\delta$-function allows to suppress the integral on $\xi_k$. The process is repeated for all values of
$k=1,\ldots,N_\text{I}-1$. Some care has to be taken with the prefactor of~$\xi_k$ in the argument of the
$\delta$-function, which will come as a denominator in front of the expression, and with the behavior of the
path~$\xi(t\pri)$ and its derivatives at the boundaries. One already sees that the result of the whole process,
in the limit $N_\text{I}\to\infty$, is that the path~$\xi(t\pri)$ will be constrained to follow the solution~$y(t\pri)$
of the differential equation
\begin{equation}\label{Eq:EqnDiffy}
-M\ddot{y}(t\pri)+\eta\dot{y}(t\pri)=\rho(t\pri)-\rho\pri(t\pri)
\end{equation}
with boundary conditions
\begin{align}\label{Eq:BoundCondy}
y(t_0)&=\xi_i,&y(t)&=\xi_f.
\end{align}
After having performed all integrals and taken the limit $N_\text{I}\to\infty$, we obtain
\begin{multline}\label{Eq:ResPF}
G(q_f,q_f\pri,q_i,q_i\pri,t)=\frac{\eta}{2\pi\hbar(1-\varepsilon)}\boxed{\sum}\exp\biggl\lbrace-S_{\text{R}}[y]-\I S_{\text{I}}[y]+\frac{\I F}{\hbar}\int_{t_0}^t\d t\pri y(t\pri)\\
+\frac{\I
M}{\hbar}\left[\dot{y}(t)\chi_f-\dot{y}(t_0)\chi_i\right]+\frac{\I\eta}{\hbar}\left[-\xi_f\chi_f+\xi_i\chi_i+\frac{1}{2}\xi_i^2\right]\biggr\rbrace,
\end{multline}
where we have introduced the notation $\varepsilon=\e^{-\gamma(t-t_0)}$. The solution of the differential
equation~(\ref{Eq:EqnDiffy}) with boundary conditions~(\ref{Eq:BoundCondy}) can be written as
$y(t\pri)=y_{\text{hom}}(t\pri)+y_{\text{part}}(t\pri)$, in terms of a solution of the associated homogeneous
differential equation
\begin{equation}
y_{\text{hom}}(t\pri)=\frac{\xi_i}{1-\varepsilon}\left[1-e^{-\gamma(t-t\pri)}\right]+\frac{\xi_f\varepsilon}{1-\varepsilon}\left[e^{\gamma(t\pri-t_0)}-1\right]
\end{equation}
and a particular solution of the differential equation
\begin{equation}
\begin{split}
y_{\text{part}}(t\pri)=&\tilde{L}\sum_{j=1}^n\sigma_j\left[\theta(t\pri-t_j)\left[1-e^{\gamma(t\pri-t_j)}\right]+\frac{[\varepsilon-e^{-\gamma(t_j-t_0)}][1-e^{\gamma(t\pri-t_0)}]}{1-\varepsilon}\right]\\
&-\tilde{L}\sum_{j\pri=1}^{n\pri}\sigma\pri_{j\pri}\left[\theta(t\pri-t\pri_{j\pri})\left[1-e^{\gamma(t\pri-t\pri_{j\pri})}\right]+\frac{[\varepsilon-e^{-\gamma(t\pri_{j\pri}-t_0)}][1-e^{\gamma(t\pri-t_0)}]}{1-\varepsilon}\right].
\end{split}
\end{equation}
The periodicity length of the tight-binding model~$\tilde{L}=2\pi\hbar/\eta L$ comes
into play at this stage.
\subsection{Evaluation of the generating function}
\begin{figure}
\begin{center}
\includegraphics{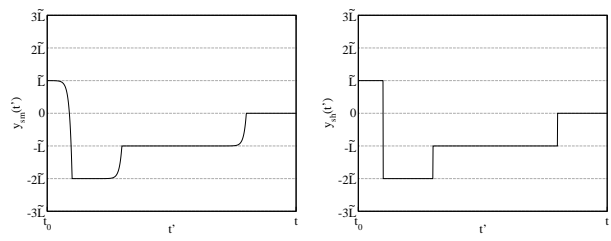}
\end{center}
\caption{Typical {\em smeared} path $y_{\text{sm}}(t\pri)$ (left) and {\em sharp} path $y_{\text{sh}}(t\pri)$
(right), as defined in Eq.~(\ref{Eq:Defysm}), respectively above Eq.~(\ref{Eq:DefTBPaths}). The height of the
steps are multiples of~$\tilde{L}$ and the edges of~$y_{\text{sm}}(t\pri)$ are smeared on a
scale~$1/\gamma$.}\label{Fig:smshpaths}
\end{figure}

The result (\ref{Eq:ResPF}) for the propagating function can now be used to obtain the generating function. One
has first to evaluate the integral~$\int\d q\e^{\lambda q}G(q,q,q_i,q_i\pri,t)$. Again, it will yield a
$\delta$-function for~$\xi_i$
\begin{equation}
\int\d q\exp\left\lbrace q\left[\frac{\I M}{\hbar}\dot{y}(t)+\lambda\right]\right\rbrace
=\frac{2\pi\hbar(1-\varepsilon)}{\eta}\delta\left(\xi_i-\Delta\xi_{\text{e}}+\Delta\xi+\frac{i\hbar(1-\varepsilon)}{\eta}\lambda\right),
\end{equation}
with~$\Delta\xi$ defined in~(\ref{Eq:DefDeltaxichi}) and
$\Delta\xi_{\text{e}}=\tilde{L}\left[\sum_{j=1}^n\sigma_j\e^{-\gamma(t_j-t_0)}-\sum_{j\pri=1}^{n\pri}\sigma\pri_{j\pri}\e^{-\gamma(t\pri_{j\pri}-t_0)}\right]$.
The generating function reads then, after transforming the variables~$q_i$ and $q_i\pri$ of the initial
integrals into~$\xi_i$ and $\chi_i$,
\begin{multline}\label{Eq:GFxichi}
\tilde{P}(\lambda,t)=\boxed{\sum}\int\d\xi_i\delta\left(\xi_i-\Delta\xi_{\text{e}}+\Delta\xi+\frac{\I\hbar(1-\varepsilon)}{\eta}\lambda\right)
\exp\biggl\lbrace-S_{\text{R}}[y]-\I S_{\text{I}}[y]+\frac{\I F}{\hbar}\int_{t_0}^t\d t\pri y(t\pri)+\frac{\I\eta}{2\hbar}\xi_i^2\biggr\rbrace\\
\times\int\d\chi_i\langle\chi_i+\xi_i/2|\hat{\rho}(t_0)|\chi_i-\xi_i/2\rangle\exp\biggl\lbrace-\frac{\I
M}{\hbar}\dot{y}(t_0)\chi_i+\frac{\I\eta}{\hbar}\xi_i\chi_i\biggr\rbrace.
\end{multline}
In the position representation, an integral of the form~$\int\d\chi_i\langle\chi_i|\cdot|\chi_i\rangle$ is a
trace~$\Tr_\text{R}\lbrace\cdot\rbrace$. With the help of the eigenstates of the momentum operator, which are
described by the wave-function $\langle q|p\rangle=\e^{\I qp/\hbar}/\sqrt{2\pi\hbar}$, one can rewrite after
some algebra
\begin{equation}
\int\d\chi_i\langle\chi_i+\xi_i/2|\hat{\rho}(t_0)|\chi_i-\xi_i/2\rangle\exp\biggl\lbrace-\frac{\I
M}{\hbar}\dot{y}(t_0)\chi_i+\frac{\I\eta}{\hbar}\xi_i\chi_i\biggr\rbrace
=\Tr\nolimits_\text{R}\left\lbrace\hat{\rho}(t_0)\exp\left[\frac{-\I
M\dot{y}(t_0)+\I\eta\xi_i}{\hbar}\hat{q}+\frac{\I\xi_i}{\hbar}\hat{p}\right]\right\rbrace.
\end{equation}
In Eq.~(\ref{Eq:GFxichi}), the integral on~$\xi_i$ can be removed, provided that one substitutes~$\xi_i$
everywhere in the integrand by the value given in the argument of the $\delta$-function. In the path~$y(t\pri)$,
this substitution yields
$\left.y(t\pri)\right|_{\xi_i=-\Delta\xi+\Delta\xi_{\text{e}}-\I\hbar(1-\varepsilon)\lambda/\eta}=y_{\text{sm},\lambda}(t\pri)$,
with
\begin{equation}
y_{\text{sm},\lambda}(t\pri)=-\frac{\I\hbar\lambda}{\eta}\left[1-\e^{-\gamma(t-t\pri)}\right]+y_{\text{sm}}(t\pri).
\end{equation}
The path $y_{\text{sm}}(t\pri)=y_{\text{sm},\lambda=0}(t\pri)$ assumes the expression
\begin{equation}\label{Eq:Defysm}
y_{\text{sm}}(t\pri)=
\tilde{L}\sum_{j=1}^n\sigma_j\left[\theta(t\pri-t_j)-1\right]\left[1-\e^{\gamma(t\pri-t_j)}\right]\\
-\tilde{L}\sum_{j\pri=1}^{n\pri}\sigma\pri_{j\pri}\left[\theta(t\pri-t\pri_{j\pri})-1\right]\left[1-\e^{\gamma(t\pri-t\pri_{j\pri})}\right].
\end{equation}
The index {\em sm} stands for {\em smeared}, because $y_{\text{sm}}(t\pri)$ is a step-like path whose edges are
smeared on a scale $1/\gamma$, as one can see on an example drawn in Fig.~\ref{Fig:smshpaths}. Collecting everything, one obtains the intermediate result
\begin{multline}\label{Eq:GFsm}
\tilde{P}(\lambda,t)=\boxed{\sum}\Tr\nolimits_\text{R}\left\lbrace\hat{\rho}(t_0)\exp\left[\left(-\frac{\I\eta}{\hbar}\Delta\xi+\lambda\right)\hat{q}+\left(-\frac{\I\eta}{\hbar}(\Delta\xi-\Delta\xi_{\text{e}})+(1-\varepsilon)\lambda\right)\frac{\hat{p}}{\eta}\right]\right\rbrace\\
\times\exp\left\lbrace-S_{\text{R}}[y_{\text{sm},\lambda}]-\I S_{\text{I}}[y_{\text{sm},\lambda}]+\frac{\I
F}{\hbar}\int_{t_0}^t\d t\pri
y_{\text{sm},\lambda}(t\pri)+\frac{\I\eta}{2\hbar}\left[-\Delta\xi+\Delta\xi_{\text{e}}-\frac{\I\hbar(1-\varepsilon)}{\eta}\lambda\right]^2\right\rbrace.
\end{multline}

\subsection{Identification with a tight-binding expression}
The following task is to rewrite the generating function~(\ref{Eq:GFsm}) in terms of the sharp
path~$y_{\text{sh}}(t\pri)$ defined above Eq.~(\ref{Eq:DefTBPaths}) and represented in Fig.~\ref{Fig:smshpaths}.
The easiest way to understand the mechanism of this transformation is to work in Fourier representation.
Technically, because these paths are defined in the time interval~$[t_0,t]$ only, one has first to continue them
in the whole time axis by defining $\bar{y}(t\pri)=[\theta(t\pri-t_0)-\theta(t\pri-t)]y(t\pri)$, in order to be
able to use the usual Fourier transform $\tilde{y}(\omega)=\int_{-\infty}^{\infty}\d
t\pri\bar{y}(t\pri)\e^{-\I\omega(t\pri-t_0)}$ and take full advantage of the usual differentiation and convolution
properties. One can then demonstrate the relation
\begin{equation}\label{Eq:RelsmshFT}
\tilde{y}_{\text{sm},\lambda}(\omega)=\frac{\I\gamma}{\omega+\I\gamma}\left[\tilde{y}_{\text{sh},\lambda}(\omega)+\frac{\Delta\xi-\Delta\xi_{\text{e}}}{\gamma}\right],
\end{equation}
with the definition
\begin{equation}\label{Eq:Defyshlambda}
\bar{y}_{\text{sh},\lambda}(t\pri)=-\frac{\I\hbar\lambda}{\eta}\left[\theta(t\pri-t_0)-\theta(t\pri-t)-\delta(t\pri-t_0)\frac{1-\varepsilon}{\gamma}\right]+\bar{y}_{\text{sh}}(t\pri).
\end{equation}
This can be done by considering the differential equation~(\ref{Eq:EqnDiffy}) in Fourier representation. This
relation means that, up to a boundary term, the Fourier transform of the smeared and sharp paths are related by
a factor~$\I\gamma/(\omega+\I\gamma)$. Let us also write the real part of the influence
phase~$S_\text{R}[y_\text{sm}]$ in Fourier representation
\begin{equation}
S_{\text{R}}[y_\text{sm}]=\frac{1}{2\pi\hbar}\int_{0}^{\infty}\d\omega
J(\omega)\coth\left(\hbar\omega\beta/2\right)\tilde{y}_\text{sm}(\omega)\tilde{y}_\text{sm}(-\omega).
\end{equation}
It can be rewritten in terms of~$\tilde{y}_\text{sh}(\omega)$ by reabsorbing the factors in a redefinition of
the spectral density as in (\ref{Eq:DefSDTB}). This will additionally yield two boundary terms.

In order to rewrite the other terms of~(\ref{Eq:GFsm}) in terms of the sharp path, it is more convenient to
rewrite relation~(\ref{Eq:RelsmshFT}) in time domain
\begin{equation}
\bar{y}_{\text{sm},\lambda}(t\pri)=\int_{t\pri}^{\infty}\d
t\pripri\e^{\gamma(t\pri-t\pripri)}\left[\gamma\bar{y}_{\text{sh},\lambda}(t\pripri)+(\Delta\xi-\Delta\xi_{\text{e}})\delta(t\pripri-t_0)\right].
\end{equation}
Collecting all terms, we get the still exact result:
\begin{multline}
\tilde{P}(\lambda,t)=\boxed{\sum}\Tr\nolimits_\text{R}\left\lbrace\hat{\rho}(t_0)\exp\left[\left(-\frac{\I\eta}{\hbar}\Delta\xi+\lambda\right)\hat{q}+\left(-\frac{\I\eta}{\hbar}(\Delta\xi-\Delta\xi_{\text{e}})+(1-\varepsilon)\lambda\right)\frac{\hat{p}}{\eta}\right]\right\rbrace\\
\times\exp\biggl\lbrace\Phi_{\text{FV}}^{\text{TB}}[y_{\text{sh},\lambda},x_{\text{sh}},t]+\frac{\I
F}{\hbar}\int_{-\infty}^\infty\d t\pri
\bar{y}_{\text{sh},\lambda}(t\pri)-\frac{\Delta\xi-\Delta\xi_{\text{e}}}{\hbar\gamma}\int_{-\infty}^{\infty}\d
t\pri
\bar{y}_{\text{sh},\lambda}(t\pri)L_{\text{R}}^{\text{TB}}(t\pri-t_0)\\
-\frac{(\Delta\xi-\Delta\xi_{\text{e}})^2}{2\hbar\gamma^2}L_{\text{R}}^{\text{TB}}(0)+\frac{\I
F(\Delta\xi-\Delta\xi_{\text{e}})}{\hbar\gamma}+\frac{\I\eta}{2\hbar}\left[-\Delta\xi+\Delta\xi_{\text{e}}-\frac{\I\hbar(1-\varepsilon)}{\eta}\lambda\right]^2\biggr\rbrace.
\end{multline}
The influence phase~$\Phi_{\text{FV}}^{\text{TB}}[y_{\text{sh},\lambda},x_{\text{sh}},t]$ is defined as
in~(\ref{Eq:DefPhiFV}), with the correlation functions $L^{\text{TB}}(\tau)$ and $M_\text{I}^{\text{TB}}(\tau)$
defined in terms of the new spectral density~$J_\text{TB}(\omega)$ given in~(\ref{Eq:DefSDTB}).
Technically,~$\Phi_{\text{FV}}^{\text{TB}}[y_{\text{sh},\lambda},x_{\text{sh}},t]$ can be rewritten in terms of
the extended path~$\bar{y}_{\text{sh},\lambda}(t\pri)$ by substituting everywhere
$\bar{y}_{\text{sh},\lambda}(t\pri)$ for $y_{\text{sh},\lambda}(t\pri)$ and extending the corresponding
integrals to the whole real axis. This applies to both integrals in the real part, yielding a prefactor~$1/2$,
and to the integrals on~$t\pri$ in the imaginary part. The average path~$x_\text{sh}(t\pri)$ was defined above
Eq.~(\ref{Eq:DefTBPaths}).

In order to go further, we now have to exploit the simplifications valid in the regime described at the end of
Section~\ref{Sec:RD}. The long-time limit yields $\varepsilon\ll 1$ and the rare transitions limit
$\Delta\xi_\text{e}\ll\Delta\xi$. Furthermore, one can see that~$L_{\text{R}}^{\text{TB}}(0)$ diverges when the
cutoff frequency~$\omega_{\text{c}}$ of an Ohmic spectral density~$J(\omega)=\eta\omega
e^{-\omega/\omega_{\text{c}}}$ tends to infinity, as in the strictly Ohmic case considered here. Then, in the
configuration sum, the contributions of the configurations of $\lbrace\sigma\rbrace$ and
$\lbrace\sigma\pri\rbrace$ for which the prefactor~$(\Delta\xi-\Delta\xi_{\text{e}})^2/2\hbar\gamma^2$ of
$L_{\text{R}}^{\text{TB}}(0)$ is minimal can be made as large as one will with respect to the contributions of
other configurations. Therefore, combining with the rare transitions limit, we can restrict the configuration
sum to configurations such that~$\Delta\xi=0$. This restricted sum is denoted with the boxed primed sum
introduced in~(\ref{Eq:DefPBsum}). Combining these properties, we find the much simpler expression
\begin{equation}\label{Eq:GFyshlambda}
\tilde{P}(\lambda,t)\sim\boxed{\sideset{}{^\prime}\sum}\Tr\nolimits_\text{R}\left\lbrace\hat{\rho}(t_0)\e^{\lambda\left(\hat{q}+\hat{p}/\eta\right)}\right\rbrace
\exp\left\lbrace\Phi_{\text{FV}}^{\text{TB}}[y_{\text{sh},\lambda},x_{\text{sh}},t]+\frac{\I
F}{\hbar}\int_{-\infty}^\infty\d t\pri
\bar{y}_{\text{sh},\lambda}(t\pri)-\frac{\I\hbar\lambda^2}{2\eta}\right\rbrace.
\end{equation}
This expression resembles already much more the one of a tight-binding generating function~(\ref{Eq:GFTB}).
\subsection{Extraction of the $\lambda$-dependence}\label{SubSec:ExtractLambda}
The last task in order to demonstrate~(\ref{Eq:DRGF}) is to extract explicitly the $\lambda$-dependence out of
the path~$y_{\text{sh},\lambda}(t\pri)$, using~(\ref{Eq:Defyshlambda}). For the influence phase, we find
\begin{multline}\label{Eq:ExtractlambdaPhiFV}
\Phi_{\text{FV}}^{\text{TB}}[y_{\text{sh},\lambda},x_{\text{sh}},t]\\
=\Phi_{\text{FV}}^{\text{TB}}[y_{\text{sh}},x_{\text{sh}},t]+\lambda\biggl[-\frac{\I}{\eta}\int_{t_0}^{t}\d
t\pri y_{\text{sh}}(t\pri)\dot{N}(t\pri)
-\Delta\chi+\int_{t_0}^{t}\d
t\pri\dot{x}_{\text{sh}}(t\pri)\e^{-\gamma(t-t\pri)}\biggr]+\frac{\hbar\lambda^2}{2\eta^2}\left[N(t)-\frac{1-\varepsilon}{\gamma}\dot{N}(t_0)\right],
\end{multline}
with the auxiliary function $N(t\pri)=\int_{t_0}^{t\pri}\d t\pripri\dot{N}(t\pripri)$ given by
\begin{equation}
\dot{N}(t\pri)=\int_{-\infty}^{\infty}\d
t\pripri\left[\theta(t\pripri-t_0)-\theta(t\pripri-t)-\delta(t\pripri-t_0)\frac{1-\varepsilon}{\gamma}\right]L_{\text{R}}^{\text{TB}}(t\pri-t\pripri).
\end{equation}
Using the explicit expression of the path~$y_{\text{sh}}(t\pri)$ given by~(\ref{Eq:DefTBPaths}) and the
property $y_{\text{sh}}(t_0)=-\Delta\xi=0$ satisfied by the paths over which the configuration sum runs,
one has
\begin{equation}
-\int_{t_0}^{t}\d t\pri y_{\text{sh}}(t\pri)\dot{N}(t\pri)=\tilde{L}\left[\sum\nolimits_{j=1}^n\sigma_j
N(t_j)-\sum\nolimits_{j\pri=1}^{n\pri}\sigma\pri_{j\pri}N(t\pri_{j\pri})\right].
\end{equation}
In the rare transitions limit, it is possible to show
\begin{equation}
N(t_\text{tr})\sim-\frac{2\eta k_\text{B}T}{\hbar}\left[t_\text{tr}-t_0-\frac{1}{\gamma}\right]
\end{equation}
when~$t_\text{tr}$ equals any of the transition times $t_j$, $t\pri_{j\pri}$. Therefore one may rewrite
\begin{equation}
-\int_{t_0}^{t}\d t\pri y_{\text{sh}}(t\pri)\dot{N}(t\pri)=\frac{2\eta k_\text{B}T}{\hbar}\int_{t_0}^{t}\d t\pri
y_{\text{sh}}(t\pri).
\end{equation}
Similarly, the term
\begin{equation}
\int_{t_0}^{t}\d
t\pri\dot{x}_{\text{sh}}(t\pri)\e^{-\gamma(t-t\pri)}=\tilde{L}\left[\sum\nolimits_{j=1}^n\sigma_j
\e^{-\gamma(t-t_j)}-\sum\nolimits_{j\pri=1}^{n\pri}\sigma\pri_{j\pri}\e^{-\gamma(t-t\pri_{j\pri})}\right]
\end{equation}
is negligible in the rare transitions limit. Finally, for the driving contribution we obtain
\begin{equation}
\frac{\I F}{\hbar}\int_{-\infty}^\infty\d t\pri \bar{y}_{\text{sh},\lambda}(t\pri)=\frac{\I
F}{\hbar}\int_{t_0}^t\d t\pri y_{\text{sh}}(t\pri)+\frac{\lambda
F}{\eta}\left[t-t_0-\frac{1-\varepsilon}{\gamma}\right].
\end{equation}
The last term in the square brackets may be neglected in the long-time limit. Putting these results
in~(\ref{Eq:GFyshlambda}) completes the proof of the duality relation~(\ref{Eq:DRGF}).

\subsection{The tight-binding generating function at long time}\label{SubSec:GFTBlongtime}
Here we shall prove that the solution of the generalized master equation~(\ref{Eq:GME}) behaves at long time as
the expression given in~(\ref{Eq:GFlongtime}). In order to discuss the long time behavior of the generating
function, it is convenient to introduce its Laplace transform
\begin{equation}
\mathcal{L}\tilde{P}_\text{TB}(\lambda,s)=\int_{t_0}^\infty\d t\e^{-s(t-t_0)}\tilde{P}_\text{TB}(\lambda,t).
\end{equation}
The long-time behavior is related to the behavior of the Laplace transform at small~$s$. Rewriting the
generalized master equation for the Laplace transform turns the convolution into a product
\begin{equation}
\mathcal{L}\tilde{P}_\text{TB}(\lambda,s)=\frac{1}{s}\left[\tilde{P}_\text{TB}(\lambda,t_0)+\int_{0}^\infty\d\tau\e^{-s\tau}\tilde{K}(\lambda,\tau)\mathcal{L}\tilde{P}_\text{TB}(\lambda,s)\right].
\end{equation}
This equation may easily be solved. Remembering that the initial populations
are~$P_l^\text{TB}(t_0)=\delta_{l,0}$, yielding ${\tilde{P}_\text{TB}(\lambda,t_0)=1}$, we obtain
\begin{equation}
\mathcal{L}\tilde{P}_\text{TB}(\lambda,s)=\left[s-\int_{0}^\infty\d\tau\e^{-s\tau}\tilde{K}(\lambda,\tau)\right]^{-1}.
\end{equation}
Let us compare this expression with the Laplace transform of the announced solution~(\ref{Eq:GFlongtime})
\begin{equation}
\mathcal{L}\tilde{P}_\text{TB}(\lambda,s)=\left[s-\sum_{m\ne0}\Gamma_m\left(\e^{\lambda
m\tilde{L}}-1\right)\right]^{-1}.
\end{equation}
Remembering that $\tilde{K}(\lambda,\tau)=\sum_{m=-\infty}^\infty \e^{\lambda m\tilde{L}}K_m(\tau)$, one sees
that the two expressions are equal in the limit of small~$s$ if one defines
\begin{equation}
\Gamma_m=\int_{0}^\infty\d\tau K_m(\tau)
\end{equation}
and provided the property $\Gamma_0=-\sum_{m\ne0}\Gamma_m$ is satisfied. This property describes the
conservation of the total population and can be verified on the explicit expressions for the transition
rates~$\Gamma_m$.
\subsection{Explicit expressions for the transition rates}\label{SubSec:TBrates}
We shall now discuss the explicit expression of the transition rates~$\Gamma_m$ in a dissipative tight-binding
model, characterized by the Hamiltonian~(\ref{Eq:DefHTB}) coupled to a bath of harmonic oscillators with
spectral density~$J_\text{TB}(\omega)$. Real-time path integrals techniques yield expressions for the transition
rates in terms of a pair of tight-binding paths. Their structure is closely related to the one of the generating
function~(\ref{Eq:GFTB}), in a way sketched in Section~\ref{SubSec:GFTBlongtime}. As discussed in
Section~\ref{Sec:App}, one can classify the different contributions to a given rate~$\Gamma_m$ with respect to
the number~$N$ of transitions occurring in the pair of tight-binding paths. These contributions form the $N$th
order rate denoted~$\Gamma_m^{(N)}$. The total rate is the sum of these contributions
$\Gamma_m=\sum_{N=2}^\infty\Gamma_m^{(N)}$. The different contributions to the $N$th order rates can themselves
be split in terms of the numbers~$n$, respectively~$n\pri$, of transitions happening in the forward
path~(\ref{Eq:DefTBForwardPath}), respectively backward path~(\ref{Eq:DefTBBackwardPath}). This means
$\Gamma_m^{(N)}=\sum_{n=0}^N\sum_{n\pri=0}^N\delta(n+n\pri,N)\Gamma_m^{(n,n\pri)}$.

Each contribution to~$\Gamma_m^{(n,n\pri)}$ can be parameterized by a set of charges
$\alpha_m^{(n,n\pri)}=\lbrace\sigma_1,\ldots,\sigma_n;\sigma\pri_1,\ldots,\sigma\pri_{n\pri}\rbrace$, which
satisfy the constraints $\sum_{j=1}^n\sigma_j=\sum_{j\pri=1}^{n\pri}\sigma\pri_{j\pri}=m$. The
charges~$\sigma_j$, respectively $\sigma\pri_{j\pri}$, characterize the transitions in the forward
path~(\ref{Eq:DefTBForwardPath}), respectively backward path~(\ref{Eq:DefTBBackwardPath}) In the tight-binding
model~(\ref{Eq:DefHTB}), they can take any positive, respectively negative, integer value, representing a
transition to the neighbor of corresponding order to the right, respectively to the left. For given $m$, $n$ and
$n\pri$, these contributions may be written~\cite{WeiBK99,BauPRB04}
\begin{equation}\label{Eq:Ratesigmasigmapri}
\Gamma_m^{(n,n\pri)}=\sum_{\left\lbrace\alpha_m^{(n,n\pri)}\right\rbrace}\Lambda\int_{-\infty}^\infty\d\tau\prod_{j=1}^{n-1}\left(\int_0^\infty\d\rho_j\right)\prod_{j\pri=1}^{n\pri-1}\left(\int_0^\infty\d\rho\pri_{j\pri}\right)\exp{\left\lbrace\Phi_\text{FV}+\Psi\right\rbrace}
\end{equation}
with the prefactor
\begin{equation}
\Lambda=\prod_{j=1}^n\left(\frac{-\I\Delta_{\sigma_j}}{\hbar}\right)\prod_{j\pri=1}^{n\pri}\left(\frac{\I\Delta^*_{\sigma\pri_{j\pri}}}{\hbar}\right),
\end{equation}
the influence phase
\begin{equation}
\Phi_\text{FV}=\frac{\tilde{L}^2}{\hbar}\Biggl[\sum_{k=1}^n\sum_{j=1}^{k-1}\sigma_k\sigma_jQ(t_k-t_j)+\sum_{k=1}^{n\pri}\sum_{j=1}^{k-1}\sigma\pri_k\sigma\pri_jQ^*(t\pri_k-t\pri_j)
-\sum_{k=1}^{n\pri}\sum_{j=1}^n\sigma\pri_k\sigma_jQ(t\pri_k-t_j)\Biggr],
\end{equation}
and the driving term
\begin{equation}
\Psi=\frac{\I
F\tilde{L}}{\hbar}\left[\sum_{j\pri=1}^{n\pri}\sigma\pri_{j\pri}t\pri_{j\pri}-\sum_{j=1}^{n}\sigma_j t_j\right].
\end{equation}
These last two expressions have to be rewritten in terms of the time variables $\rho_j=t_{j+1}-t_j$,
$\rho\pri_{j\pri}=t\pri_{j\pri+1}-t\pri_{j\pri}$, and
$\tau=\frac{1}{n\pri}\sum_{j\pri=1}^{n\pri}t\pri_{j\pri}-\frac{1}{n}\sum_{j=1}^nt_j$. If~$n=0$,
respectively~$n\pri=0$, there are no integrals on~$\tau$ and~$\rho\pri_{j\pri}$, respectively~$\tau$
and~$\rho_j$. The twice-integrated bath correlation function is given in~(\ref{Eq:DefTIBathCorr}).

The expression~(\ref{Eq:Ratesigmasigmapri}) is valid up to third-order~$N=n+n\pri=2,3$ only. At higher
order~$N>3$, reducible contributions have to be subtracted~\cite{SasPRB96}.

%
%
\begin{figure}
\begin{center}
\includegraphics{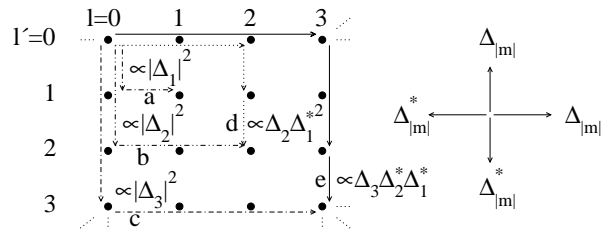}
\end{center}
\caption{Left:~Graphical representation of the pairs of tight-binding paths corresponding to some contributions
to the second-order~(a,b,~and~c) and third-order~(d~and~e) transition rates, with the corresponding dependence
on the couplings~$\Delta_m$. Right:~The couplings associated with the transitions in the forward
path~(horizontal) and backward path~(vertical).}\label{Fig:ReprPaths}
\end{figure}
The problem is to find all configurations of pairs of tight-binding paths involving $N$~transitions in total,
and satisfying the constraints $\sum_{j=1}^n\sigma_j=\sum_{j\pri=1}^{n\pri}\sigma\pri_{j\pri}=m$, or in other
words to perform the configuration sum on~$\alpha_m^{(n,n\pri)}$ in Eq.~(\ref{Eq:Ratesigmasigmapri}). A
graphical representation can help us in this task. Let us start from a two-dimensional lattice parameterized by
pairs of integers $(l,l\pri)$. Let us represent by an horizontal arrow of length~$m$ to the right a transition
in the forward tight-binding path to the neighbor of order $m$ to the right. Such a transition is characterized
by a charge~$\sigma=+m$. A transition to the left neighbor of order $m$, characterized by a charge~$\sigma=-m$,
is represented by an arrow of length~$m$ to the left. Likewise, we represent transitions happening in the
backward path, characterized by a charge~$\sigma\pri=+m$, respectively~$\sigma\pri=-m$, by vertical arrows of
length~$m$, pointing downwards, respectively upwards, in our diagram. Together, the arrows corresponding to the
transitions happening in the pair of tight-binding paths will draw a path in the diagram. With this
representation, it is now easy to find the paths satisfying the constraints
$\sum_{j=1}^n\sigma_j=\sum_{j\pri=1}^{n\pri}\sigma\pri_{j\pri}=m$. If one starts from a diagonal site of our
diagram, say at position~$(0,0)$, the constraints select those paths which end up in the diagonal site~$(m,m)$.
Some examples are given in Fig.~\ref{Fig:ReprPaths}. Let us now look in detail to the second-order and third-order contributions.

In order to understand the different contributions to the second-order rates~$\Gamma_m^{(2)}$, let us first look
at those contributions which involve only transitions to the nearest neighbors. In this case, the charges
$\sigma_j$ and $\sigma\pri_{j\pri}$ may only take the values $+1$ and $-1$, denoting transitions to the nearest
neighbor to the right, respectively to the left. These transitions are mediated by the
couplings~$\Delta_{\pm1}$. In other words, we consider the situation described by the
Hamiltonian~(\ref{Eq:DefHTB}) with~$\Delta_m=0$ for~$|m|\ge2$. All contributing configurations are listed in
Table~\ref{Table:2ndRates}. From there, it is clear that there is only one contribution to~$\Gamma_1^{(2)}$, and
only one to~$\Gamma_{-1}^{(2)}$, which read
\begin{equation}
\Gamma_{\pm1}^{(2)}=\frac{\left|\Delta_1\right|^2}{\hbar^2}\int_{-\infty}^\infty\d\tau\e^{-(\tilde{L}^2/\hbar)Q(\tau)\pm\I(F\tilde{L}/\hbar)\tau}.
\end{equation}
From the numerous contributions to~$\Gamma_0^{(2)}$, we can explicitly check the relation
$\Gamma_0^{(2)}=-\Gamma_{+1}^{(2)}-\Gamma_{-1}^{(2)}$ required in Section~\ref{SubSec:GFTBlongtime}.
\begin{table}
\begin{center}
\begin{tabular}{lcccc}
\hline\hline $\Gamma_m^{(n,n\pri)}$ & $\alpha_m^{(n,n\pri)}$ &
$\Lambda$ & $\Phi_\text{FV}$ & $\Psi$\\
&&(units of $\frac{1}{\hbar^2}$)&(units of $\frac{\tilde{L}^2}{\hbar}$)&(units of $\frac{\I
F\tilde{L}}{\hbar}$)\\\hline
$\Gamma_1^{(1,1)}$ & $\lbrace+1;+1\rbrace$ & $\left|\Delta_1\right|^2$ & $-Q(\tau)$ & $\tau$\\
$\Gamma_0^{(2,0)}$ & $\lbrace+1,-1;\varnothing\rbrace$ & $-\left|\Delta_1\right|^2$ & $-Q(\rho_1)$ & $\rho_1$\\
& $\lbrace-1,+1;\varnothing\rbrace$ & $-\left|\Delta_1\right|^2$ & $-Q(\rho_1)$ & $-\rho_1$\\
$\Gamma_0^{(0,2)}$ & $\lbrace\varnothing;+1,-1\rbrace$ & $-\left|\Delta_1\right|^2$ & $-Q^*(\rho\pri_1)$ & $-\rho\pri_1$\\
& $\lbrace\varnothing;-1,+1\rbrace$ & $-\left|\Delta_1\right|^2$ & $-Q^*(\rho\pri_1)$ & $\rho\pri_1$\\
$\Gamma_{-1}^{(1,1)}$ & $\lbrace-1;-1\rbrace$ & $\left|\Delta_1\right|^2$ & $-Q(\tau)$ & $-\tau$\\
\hline\hline
\end{tabular}
\end{center}
\caption{The different contributions to the rates~$\Gamma_m^{(n,n\pri)}$, at second order~$n+n\pri=2$, involving
only transitions to the nearest neighbors, characterized by charges $\sigma_j,\sigma\pri_{j\pri}=\pm1$, and
mediated by the coupling~$\Delta_{\pm1}$. Each contribution is specified by
$\alpha_m^{(n,n\pri)}=\lbrace\sigma_1,\ldots,\sigma_n;\sigma\pri_1,\ldots,\sigma\pri_{n\pri}\rbrace$, given in
the second column. The last three columns give the explicit expression of the prefactor, the influence phase,
and the driving term for the corresponding contribution.}\label{Table:2ndRates}
\end{table}

Let us now allow all couplings~$\Delta_m$ in the Hamiltonian~(\ref{Eq:DefHTB}) to be nonzero and ask again the
question: which pairs of tight-binding paths involving two transitions contribute to the transition rates? In
the graphical representation shown in Fig.~\ref{Fig:ReprPaths} and described above, those paths are
represented by a succession of two arrows connecting the initial diagonal site~$(0,0)$ with any final diagonal
site~$(m,m)$. It is clear that, if the first transition in any of the two paths reaches a neighbor site at
distance~$m$, the second transition must also happen between sites at the same distance~$m$, in order to end
up back in the diagonal. At second order, there will thus be no contributions to the rates combining two
different couplings $\Delta_m$ and $\Delta_{m\pri}$. One can convince oneself that all $6$~possible combinations
for a given~$m$ are similar to the ones listed in Table~\ref{Table:2ndRates}, with~$\pm1$ replaced by~$\pm m$ in
$\alpha_m^{(n,n\pri)}$. They are proportional to~$\left|\Delta_m\right|^2$ and give contributions to~$\Gamma_{\pm m}^{(2)}$ and~$\Gamma_0^{(2)}$. One easily sees that the general expression for the influence phase
is $\Phi_\text{FV}=-m^2(\tilde{L}^2/\hbar)Q(\tau)$, and for the driving term $\Psi=\pm\I
m(F\tilde{L}/\hbar)\tau$. Thus, the contributions to the second-order transition rates~$\Gamma_m^{(2)}$, for
any~$m\ne0$, can be described by the expression~(\ref{Eq:2ndOrderRate}). For~$m=0$, the relation
$\Gamma_0=-\sum_{m\ne0}\Gamma_m$ can be explicitly verified.

We shall now investigate some contributions to the third-order rates~$\Gamma_m^{(3)}$. Let us first restrict
ourselves to the case where, in the Hamiltonian~(\ref{Eq:DefHTB}), only the couplings to the nearest
neighbors~$\Delta_{\pm1}$ and next-nearest neighbors~$\Delta_{\pm2}$ are nonzero. With the help of the graphical
representation shown in Fig.~\ref{Fig:ReprPaths}, one finds $24$~contributions in the $\sigma\text{-}\sigma\pri$
description. They are listed in Table~\ref{Table:3rd112Rates}. They all involve two transitions to a nearest
neighbor and one to a next-nearest neighbor, in other words twice~$\Delta_{\pm1}$ and once~$\Delta_{\pm2}$. We
shall group them under the notation~$\Gamma_m^{(3)}[112]$ in order to distinguish them from other contributions
which will arise when we shall switch on higher-order couplings~$\Delta_m,\ |m|>2$.
\begin{table}
\begin{center}
\begin{tabular}{lcccc}
\hline\hline $\Gamma_m^{(n,n\pri)}$ & $\alpha_m^{(n,n\pri)}$ &
$\Lambda$ & $\Phi_\text{FV}$ & $\Psi$ \\
&&&(units of $\frac{\tilde{L}^2}{\hbar}$)&(units of $\frac{\I F\tilde{L}}{\hbar}$)\\
\hline $\Gamma_2^{(2,1)}$ & $\lbrace+1,+1;+2\rbrace$ & $\lambda_{112}$ &
$Q(\rho_1)-2Q(\tau+\frac{1}{2}\rho_1)-2Q(\tau-\frac{1}{2}\rho_1) $ & $2\tau$\\
$\Gamma_2^{(1,2)}$ & $\lbrace+2;+1,+1\rbrace$ & $\lambda_{112}^*$ &
$Q^*(\rho\pri_1)-2Q(\tau+\frac{1}{2}\rho\pri_1)-2Q(\tau-\frac{1}{2}\rho\pri_1) $ & $2\tau$\\
$\Gamma_1^{(2,1)}$ & $\lbrace-1,+2;+1\rbrace$ & $-\lambda_{112}^*$ &
$-2Q(\rho_1)+Q(\tau+\frac{1}{2}\rho_1)-2Q(\tau-\frac{1}{2}\rho_1) $ & $\tau-\frac{3}{2}\rho_1$\\
& $\lbrace+2,-1;+1\rbrace$ & $-\lambda_{112}^*$ &
$-2Q(\rho_1)-2Q(\tau+\frac{1}{2}\rho_1)+Q(\tau-\frac{1}{2}\rho_1) $ & $\tau+\frac{3}{2}\rho_1$\\
$\Gamma_1^{(1,2)}$ & $\lbrace+1;-1,+2\rbrace$ & $-\lambda_{112}$ &
$-2Q^*(\rho\pri_1)-2Q(\tau+\frac{1}{2}\rho\pri_1)+Q(\tau-\frac{1}{2}\rho\pri_1) $ & $\tau+\frac{3}{2}\rho\pri_1$\\
& $\lbrace+1;+2,-1\rbrace$ & $-\lambda_{112}$ &
$-2Q^*(\rho\pri_1)+Q(\tau+\frac{1}{2}\rho\pri_1)-2Q(\tau-\frac{1}{2}\rho\pri_1) $ & $\tau-\frac{3}{2}\rho\pri_1$\\
$\Gamma_0^{(3,0)}$ & $\lbrace+1,+1,-2;\varnothing\rbrace$ & $-\lambda_{112}$ &
$Q(\rho_1)-2Q(\rho_1+\rho_2)-2Q(\rho_2) $ & $\rho_1+2\rho_2$\\
& $\lbrace+1,-2,+1;\varnothing\rbrace$ & $-\lambda_{112}$ &
$-2Q(\rho_1)+Q(\rho_1+\rho_2)-2Q(\rho_2) $ & $\rho_1-\rho_2$\\
& $\lbrace-2,+1,+1;\varnothing\rbrace$ & $-\lambda_{112}$ &
$-2Q(\rho_1)-2Q(\rho_1+\rho_2)+Q(\rho_2) $ & $-2\rho_1-\rho_2$\\
 & $\lbrace-1,-1,+2;\varnothing\rbrace$ & $\lambda_{112}^*$ &
$Q(\rho_1)-2Q(\rho_1+\rho_2)-2Q(\rho_2) $ & $-\rho_1-2\rho_2$\\
& $\lbrace-1,+2,-1;\varnothing\rbrace$ & $\lambda_{112}^*$ &
$-2Q(\rho_1)+Q(\rho_1+\rho_2)-2Q(\rho_2) $ & $-\rho_1+\rho_2$\\
& $\lbrace+2,-1,-1;\varnothing\rbrace$ & $\lambda_{112}^*$ &
$-2Q(\rho_1)-2Q(\rho_1+\rho_2)+Q(\rho_2) $ & $+2\rho_1+\rho_2$\\
$\Gamma_0^{(0,3)}$ & $\lbrace\varnothing;+1,+1,-2\rbrace$ & $-\lambda_{112}^*$ &
$Q^*(\rho\pri_1)-2Q^*(\rho\pri_1+\rho\pri_2)-2Q^*(\rho\pri_2) $ & $-\rho\pri_1-2\rho\pri_2$\\
& $\lbrace\varnothing;+1,-2,+1\rbrace$ & $-\lambda_{112}^*$ &
$-2Q^*(\rho\pri_1)+Q^*(\rho\pri_1+\rho\pri_2)-2Q^*(\rho\pri_2) $ & $-\rho\pri_1+\rho\pri_2$\\
& $\lbrace\varnothing;-2,+1,+1\rbrace$ & $-\lambda_{112}^*$ &
$-2Q^*(\rho\pri_1)-2Q^*(\rho\pri_1+\rho\pri_2)+Q^*(\rho\pri_2) $ & $2\rho\pri_1+\rho\pri_2$\\
& $\lbrace\varnothing;-1,-1,+2\rbrace$ & $\lambda_{112}$ &
$Q^*(\rho\pri_1)-2Q^*(\rho\pri_1+\rho\pri_2)-2Q^*(\rho\pri_2) $ & $\rho\pri_1+2\rho\pri_2$\\
& $\lbrace\varnothing;-1,+2,-1\rbrace$ & $\lambda_{112}$ &
$-2Q^*(\rho\pri_1)+Q^*(\rho\pri_1+\rho\pri_2)-2Q^*(\rho\pri_2) $ & $\rho\pri_1-\rho\pri_2$\\
& $\lbrace\varnothing;+2,-1,-1\rbrace$ & $\lambda_{112}$ &
$-2Q^*(\rho\pri_1)-2Q^*(\rho\pri_1+\rho\pri_2)+Q^*(\rho\pri_2) $ & $-2\rho\pri_1-\rho\pri_2$\\
$\Gamma_{-1}^{(2,1)}$ & $\lbrace+1,-2;-1\rbrace$ & $\lambda_{112}$ &
$-2Q(\rho_1)+Q(\tau+\frac{1}{2}\rho_1)-2Q(\tau-\frac{1}{2}\rho_1) $ & $-\tau+\frac{3}{2}\rho_1$\\
& $\lbrace-2,+1;-1\rbrace$ & $\lambda_{112}$ &
$-2Q(\rho_1)-2Q(\tau+\frac{1}{2}\rho_1)+Q(\tau-\frac{1}{2}\rho_1) $ & $-\tau-\frac{3}{2}\rho_1$\\
$\Gamma_{-1}^{(1,2)}$ & $\lbrace-1;+1,-2\rbrace$ & $\lambda_{112}^*$ &
$-2Q^*(\rho\pri_1)-2Q(\tau+\frac{1}{2}\rho\pri_1)+Q(\tau-\frac{1}{2}\rho\pri_1) $ & $-\tau-\frac{3}{2}\rho\pri_1$\\
& $\lbrace-1;-2,+1\rbrace$ & $\lambda_{112}^*$ &
$-2Q^*(\rho\pri_1)+Q(\tau+\frac{1}{2}\rho\pri_1)-2Q(\tau-\frac{1}{2}\rho\pri_1) $ & $-\tau+\frac{3}{2}\rho\pri_1$\\
$\Gamma_{-2}^{(2,1)}$ & $\lbrace-1,-1;-2\rbrace$ & $-\lambda_{112}^*$ &
$Q(\rho_1)-2Q(\tau+\frac{1}{2}\rho_1)-2Q(\tau-\frac{1}{2}\rho_1) $ & $-2\tau$\\
$\Gamma_{-2}^{(1,2)}$ & $\lbrace-2;-1,-1\rbrace$ & $-\lambda_{112}$ &
$Q^*(\rho\pri_1)-2Q(\tau+\frac{1}{2}\rho\pri_1)-2Q(\tau-\frac{1}{2}\rho\pri_1) $ & $-2\tau$\\
\hline\hline
\end{tabular}
\end{center}
\caption{The different contributions to the rates~$\Gamma_m^{(n,n\pri)}$, at third order~$n+n\pri=3$, involving
only transitions to the nearest neighbors and next-nearest neighbors, characterized by charges
$\sigma_j,\sigma\pri_{j\pri}=\pm1,\pm2$, and mediated by the couplings $\Delta_{\pm1}$ and $\Delta_{\pm2}$. Each
contribution is specified by
$\alpha_m^{(n,n\pri)}=\lbrace\sigma_1,\ldots,\sigma_n;\sigma\pri_1,\ldots,\sigma\pri_{n\pri}\rbrace$, given in
the second column. The third column gives the explicit expression of the prefactor, in terms
of~$\lambda_{112}=-\I\Delta_1^2\Delta_2^*/\hbar^3$, for the corresponding contribution. The last two columns
give the influence phase and the driving term.}\label{Table:3rd112Rates}
\end{table}

Let us first look at~$\Gamma_2^{(3)}[112]$. Changing the integration variable variable~$\tau$ into $-\tau$ and
using the property $Q(-\tau)=Q^*(\tau)$, one can see that the second contribution~$\Gamma_2^{(1,2)}$ is the
complex conjugate of the first one~$\Gamma_2^{(2,1)}$. Put together, they can be rewritten
\begin{equation}
\Gamma_2^{(3)}[112]=\frac{2}{\hbar^3}\Im\left\lbrace\Delta_1^2\Delta_2^*\int_{-\infty}^\infty\d\tau\int_0^\infty\d\rho\e^{\frac{\tilde{L}^2}{\hbar}\left[Q(\rho)-2Q(\tau+\frac{1}{2}\rho)-2Q(\tau-\frac{1}{2}\rho)\right]+2\I\frac{F\tilde{L}}{\hbar}\tau}\right\rbrace.
\end{equation}
Likewise, we obtain
\begin{equation}
\Gamma_{-2}^{(3)}[112]=\frac{2}{\hbar^3}\Im\left\lbrace{\Delta_1^*}^2\Delta_2\int_{-\infty}^\infty\d\tau\int_0^\infty\d\rho\e^{\frac{\tilde{L}^2}{\hbar}\left[Q(\rho)-2Q(\tau+\frac{1}{2}\rho)-2Q(\tau-\frac{1}{2}\rho)\right]-2\I\frac{F\tilde{L}}{\hbar}\tau}\right\rbrace.
\end{equation}

Following the same procedure, one can show that the contributions to~$\Gamma_1^{(2,1)}$ are the complex
conjugate of the ones to~$\Gamma_1^{(1,2)}$, yielding
\begin{multline}
\Gamma_1^{(3)}[112]=-\frac{2}{\hbar^3}\Im\biggl\lbrace\Delta_1^2\Delta_2^*\int_{-\infty}^\infty\d\tau\int_0^\infty\d\rho
\Bigl[\e^{\frac{\tilde{L}^2}{\hbar}\left[-2Q(-\rho)-2Q(\tau+\frac{1}{2}\rho)+Q(\tau-\frac{1}{2}\rho)\right]+\I\frac{F\tilde{L}}{\hbar}(\tau+\frac{3}{2}\rho)}\\
+\e^{\frac{\tilde{L}^2}{\hbar}\left[-2Q(-\rho)+Q(\tau+\frac{1}{2}\rho)-2Q(\tau-\frac{1}{2}\rho)\right]+\I\frac{F\tilde{L}}{\hbar}(\tau-\frac{3}{2}\rho)}\Bigr]\biggr\rbrace.
\end{multline}
Both parts of the expression can be partially merged by substituting~$\tau-\frac{3}{2}\rho$ for~$\tau$ in the
first one, and~$\tau+\frac{3}{2}\rho$ for~$\tau$ in the second one. This yields
\begin{equation}
\Gamma_1^{(3)}[112]=-\frac{2}{\hbar^3}\Im\biggl\lbrace\Delta_1^2\Delta_2^*\int_{-\infty}^\infty\d\tau\int_0^\infty\d\rho\e^{-\frac{2\tilde{L}^2}{\hbar}Q(-\rho)+\I\frac{F\tilde{L}}{\hbar}\tau}
\Bigl[\e^{\frac{\tilde{L}^2}{\hbar}\left[-2Q(\tau+\rho)+Q(\tau+2\rho)\right]}+\e^{\frac{\tilde{L}^2}{\hbar}\left[-2Q(\tau-\rho)+Q(\tau-2\rho)\right]}\Bigr]\biggr\rbrace.
\end{equation}
A similar expression is obtained for
\begin{equation}
\Gamma_{-1}^{(3)}[112]=-\frac{2}{\hbar^3}\Im\biggl\lbrace{\Delta_1^*}^2\Delta_2\int_{-\infty}^\infty\d\tau\int_0^\infty\d\rho\e^{-\frac{2\tilde{L}^2}{\hbar}Q(-\rho)-\I\frac{F\tilde{L}}{\hbar}\tau}
\Bigl[\e^{\frac{\tilde{L}^2}{\hbar}\left[-2Q(\tau+\rho)+Q(\tau+2\rho)\right]}+\e^{\frac{\tilde{L}^2}{\hbar}\left[-2Q(\tau-\rho)+Q(\tau-2\rho)\right]}\Bigr]\biggr\rbrace.
\end{equation}

Upon using the modulus and phase of the couplings~$\Delta_m=\left|\Delta_m\right|\e^{\I\varphi_m}$, all these
rates may be rewritten in the common expression~(\ref{Eq:3rdOrder112Rate}).

The rate~$\Gamma_0^{(3)}[112]$ here again satisfies $\Gamma_0^{(3)}[112]=-\sum_{m\ne0}\Gamma_m^{(3)}[112]$. This
relation can be explicitly checked with the expressions given in Table~\ref{Table:3rd112Rates}. One has to take
into account the various chronological orderings of the transition times in the forward and backward paths, which
lead to $48$~configurations. By proper substitutions in the time integrals, one can then check that they
compensate each other two by two.

Let us now allow the couplings~$\Delta_{\pm3}$ to be nonzero. Still using the graphical representation shown in
Fig.~\ref{Fig:ReprPaths}, one finds $48$~additional contributions to the third-order rates. They all involve
one transition of each of the three different kinds, therefore we group them under the
notation~$\Gamma_m^{(3)}[123]$. Applying the same procedure as for the third-order contributions involving only
$\Delta_{\pm1}$ and $\Delta_{\pm2}$, we arrive at the expression~(\ref{Eq:3rdOrder123Rate}).

The generalization to higher-range couplings~$\Delta_m,\ |m|>3$, and larger number of transitions now becomes
clear. If we would allow the couplings~$\Delta_{\pm4}$ to be nonzero as well, we would get additional
third-order contributions involving~$\Delta_{\pm4}$ with $\Delta_{\pm3}$ and $\Delta_{\pm1}$, as well as
contributions involving~$\Delta_{\pm4}$ and twice~$\Delta_{\pm2}$. In order to write down the explicit
expressions for these contributions, one has to follow a procedure similar to what we did here above. Going to
fourth or higher order~$N$, the procedure would also remain the same. One can see that the number of
contributions increases factorially with the number of transitions. Starting from fourth order, one has also to
subtract the reducible contributions in order to evaluate the transition rates.

\section{Conclusion}
We have developed a method yielding the duality relations (\ref{Eq:DRGF}) and (\ref{Eq:DRPos}) between the
long-time dynamics in a tilted ratchet potential in the presence of dissipation, and the long-time dynamics in a
driven dissipative tight-binding model. Detailed proofs have been presented. The formalism has been applied to
the evaluation of the current in quantum ratchet systems, yielding an expression~(\ref{Eq:vR}) in terms of the
transition rates in the tight-binding system. Consequently, the evaluation of these rates has been discussed in
Section~\ref{Sec:App}. In particular, the obtained expressions show the explicit dependence of the ratchet
current on the parameters of the ratchet potential.

This approach allows to investigate quantum ratchet systems in the weak dissipation limit, which is beyond the
validity range of many of the other theoretical approaches. With respect to the perturbative approach of
Ref.~\cite{SchPRB02}, the method reported here has the advantage that the nonlinear regime of a large driving
force, eventually leading into the classical regime, can be reached. On the other hand, the continuous system
considered is essentially different from the tight-binding molecular wire investigated in Ref.~\cite{LehPRL02},
and may thus apply to different experimental situations. Weak dissipation is even a favorable situation in our
approach. Indeed, the duality relation links a situation of weak dissipation in the original model with strong
dissipation in the dual tight-binding model, and vice versa. Therefore, one is brought to evaluate the
tight-binding transition rates in the limit of strong dissipation, where it suffices to consider the lowest
orders in the tunneling amplitude.

\section*{Acknowledgments}We enjoyed fruitful discussions with U.~Weiss. This work was supported by the Dutch Foundation for Fundamental Research on Matter (FOM).

\end{document}